\documentclass[12pt]{article}
\usepackage{eucal}

\def\PL #1 #2 #3 {{\rm Phys. Lett.} {\bf#1} (#3) #2}
\def\NP #1 #2 #3 {{\rm Nucl. Phys.} {\bf#1} (#3) #2}
\def\ZP #1 #2 #3 {{\rm Z. Phys.} {\bf#1} (#3) #2}
\def\PRL #1 #2 #3 {{\rm Phys. Rev. Lett.} {\bf #1} (#3) #2}
\def\PR #1 #2 #3 {{\rm Phys. Rev.} {\bf#1} (#3) #2}
\def\MPL #1 #2 #3 {{\rm Mod. Phys. Lett.} {\bf#1} (#3) #2}
\def\RMP #1 #2 #3 {{\rm Rev.~Mod. Phys.} {\bf#1} (#3) #2}
\def\IJMP #1 #2 #3 {{\rm Int. J.~Mod. Phys.} {\bf#1} (#3) #2}
\def\ifm{\ifmmode}

\def\als{\ifm \alpha_s \else $\alpha_s$\fi}
\def\go{\ifm \rightarrow \else $\rightarrow$\fi}

\def\eps{\ifm \epsilon \else $\epsilon $\fi}
\def\fac{
(\frac{N}{2\pi}) \frac{1}{\Gamma (1-\epsilon )}
\Big(\frac{4\pi \mu^2}{s_{\rm min}}\Big)^\epsilon \frac{1}{\eps }}
\def\facm{(\frac{N}{2\pi})}
\def\facml{\left(\frac{N}{2\pi}\right)\left({4\pi\mu^2\over\mu_D^2} \right)^\epsilon
  {1\over \Gamma(1-\epsilon)}{1\over\epsilon} }

\def\msb{\ifm \overline{\rm MS}\, \else $\overline{\rm MS}\, $\fi}
\def\mmsb{\ifm \overline{\rm MMS}\, \else $\overline{\rm MMS}\, $\fi}
\def\ms{\ifm {\rm MS}\, \else $\rm MS\, $\fi}
\def\Q{\ifm {\cal Q}\, \else ${\cal Q}\,$\fi}
\def\H{\ifm {\cal H}\, \else ${\cal H}\,$\fi}
\def\cdh{\ifm {\cal D}_h^\H \else $ {\cal D}_h^\H$\fi}
\def\dh{\ifm D_h^\H \else $D_h^\H$\fi}
\def\ch{\ifm T_{h}^\H \else $T_{h}^\H$\fi}
\def\ng{\ifm n_g \else $n_g$\fi}

\newcommand{\notp}{\ \hbox{{$p$}\kern-.43em\hbox{/}}}
\newcommand{\notE}{\ \hbox{{$E$}\kern-.43em\hbox{/}}}
\newcommand{\beq}{\begin{equation}}
\newcommand{\eeq}{\end{equation}}
\newcommand{\beqn}{\begin{eqnarray}}
\newcommand{\eeqn}{\end{eqnarray}}
\newcommand{\beqs}{\begin{eqnarray*}}
\newcommand{\eeqs}{\end{eqnarray*}}
\newcommand{\half}{\textstyle{1\over 2}}

\def\gtap{\raisebox{-.4ex}{\rlap{$\sim$}} \raisebox{.4ex}{$>$}}   

\def\rp{\rm p}
\def\rjets{{\rm jets}}
\def\rpartons{{\rm partons}}
\def\smin{s_{\rm min}}

\def\ts2p3{s_{2'3}}

\catcode`\@=11
\def\section{\@startsection{section}{1}{\z@}{3.5ex plus 1ex minus .2ex}
{2.3ex plus .2ex}{\large\bf}}
\def\thesection{\arabic{section}.}

\def\appendix{\setcounter{section}{0}
 \def\thesection{Appendix \Alph{section}:}
 \def\theequation{\Alph{section}.\arabic{equation}}}
 
\def\@citex[#1]#2{\if@filesw\immediate\write\@auxout{\string\citation{#2}}\fi
  \def\@citea{}\@cite{\@for\@citeb:=#2\do
    {\@citea\def\@citea{,\penalty\@m}\@ifundefined
       {b@\@citeb}{{\bf ?}\@warning
       {Citation `\@citeb' on page \thepage \space undefined}}%
\hbox{\csname b@\@citeb\endcsname}}}{#1}}

\def\citer{\@ifnextchar [{\@tempswatrue\@citexr}{\@tempswafalse\@citexr[]}}
\def\@citexr[#1]#2{\if@filesw\immediate\write\@auxout{\string\citation{#2}}\fi
  \def\@citea{}\@cite{\@for\@citeb:=#2\do
    {\@citea\def\@citea{--\penalty\@m}\@ifundefined
       {b@\@citeb}{{\bf ?}\@warning
       {Citation `\@citeb' on page \thepage \space undefined}}%
\hbox{\csname b@\@citeb\endcsname}}}{#1}}
\relax

 1
 1
 1

\begin{document}
\thispagestyle{empty}
\begin{flushright}
\hfill{OITS-663\\}
\hfill{NIKHEF-98-010\\}
\hfill{hep-ph/9812415}
\end{flushright}
\vskip 2cm
\begin{center}
{\bf\Large\sc Next-To-Leading Order Cross Sections} 
\vglue .2cm
{\bf\Large\sc for Tagged Reactions}
\vglue .4cm
\vglue 1.4cm
\begin{sc}
St\'{e}phane Keller
\vglue 0.2cm
\end{sc}
{\it Institute of Theoretical Science\\
 University of Oregon, Eugene, OR 97408, USA}
\vglue 0.5cm
\begin{sc}
 Eric Laenen\\
\vglue 0.4cm
\end{sc}
{\it NIKHEF Theory Group\\
P.O. Box 41882, 1009 DB Amsterdam, The Netherlands}
\end{center}

\vglue 1.5cm
\begin{abstract}
\par \vskip .1in \noindent We extend the phase space slicing method of
Giele, Glover and Kosower for performing next-to-leading order jet
cross section calculations in two important ways: we show how to
include fragmentation functions and how to include massive
particles. These extensions allow the application of this method not
just to jet cross sections but also to cross sections in which a
particular final state particle, including a $D$ or $B$-meson, is
tagged.
\end{abstract}

\vfill

\newpage
\setcounter{page}{1}
\section{Introduction}

\noindent 

The identification (``tagging'') of one or more particular particles in
the final state of reactions at colliders has been used succesfully in
the past to enhance specific
signals.  A recent tremendous success was of course the top quark
discovery~\cite{top} at the Tevatron, where $b$ quark tagging was used
to suppress the background. 
It is expected that future analyses will increasingly rely on
tagging various kinds of particles, witness the large effort spent on
developing Cherenkov (``RICH'') and micro-vertex detectors.  Heavy
($D$ and $B$) mesons in particular will certainly continue to
play an essential part in this.  It is imperative that the theoretical
side of such analyses keep pace with these developments.  A key
ingredient of this theoretical effort must be the construction of
fully exclusive Monte Carlo type programs, which include exact higher
order QCD corrections and appropriate fragmentation functions that
parametrize the transition of a parton to the particular hadron
being tagged.

It is our goal to provide a systematic general-purpose method 
for constructing such Monte Carlo programs. In this paper we
present the formalism. Numerical studies and
applications will appear elsewhere \cite{KLnum}.

Our method is an extended version of one already employed in the
calculation of higher order jet cross sections, and is known as the
``phase space slicing method'' \cite{BOO,GG92,GGK93}. One alternative
method for calculating general jet cross sections is the 
``subtraction method'' \cite{Subtraction}. This method has also been
applied to inclusive heavy quark cross sections \cite{MNR}.
Another, quite recent method is the ``dipole
method'' \cite{CaSe}, which can also be used for tagged reactions, and
whose generalization to heavy quarks is in progress \cite{CaSeFu}.
Yet another method, employing a small gauge
boson mass, was used in \cite{Vermaseren}.
The above methods differ in how they approximate the phase space and
matrix elements in the neighborhood of divergent regions.

The particular version of the phase space slicing method we extend is
given in the papers of Giele, Glover and Kosower in Refs.~\cite{GG92}
and \cite{GGK93}.  Their version minimizes computational effort in the
calculation of jet cross sections by using the concepts of color
ordering and crossing.  The latter property, implemented in higher
orders with the help of so-called crossing functions, allows the
calculation of the matrix elements to be performed with all
partons in the final state. We will employ both concepts here, and
introduce the final state counterparts to the crossing functions,
``tagging functions''.  In addition we compute the heavy quark
contributions to the matrix elements, 
and to the crossing and tagging functions.
The results presented in this paper have already partly
been used in Ref.~\cite{GKL}, and were briefly presented in
\cite{GKL0}.

The paper is organized as follows.  In section 2 we review the basics
of the phase space slicing method in the formalism of Giele, Glover
and Kosower, and recall how it is used for computing general jet cross
sections.  In section 3 we extend their method to include
fragmentation functions. The extension to heavy quarks is given in
section 4.  We summarize and conclude in section 5.

\section{Phase Space Slicing Method for Jet Cross Sections}

\noindent
In order to be self-contained and to 
establish notation used in later sections,
we review in this section
the phase space slicing (PSS)
method, as given in 
Refs.~\cite{GG92,GGK93}, for calculating jet cross sections to next-to-leading order 
(NLO) in QCD. The method admits 
in principle an arbitrary
number of jets in the final state, and is applicable to jet cross
sections with either leptons or hadrons or both in the initial
state. It allows for the study of any distribution of a given jet
cross section, and for changes of jet-finding algorithms without
analytical recalculation of matrix elements.

Consider the situation where at lowest order one has $n$ massless
partons $\rp$ in the final state produced by a vector boson $V$, 
i.e. the reaction:
\begin{equation}
\Big(l\bar{l'} \rightarrow\Big) V \rightarrow n\,\rp.
\label{revone}
\end{equation}
$V$ itself can be the result of a lepton ($l$) anti-lepton
($\bar{l'}$) collision.  At this lowest order each parton is
associated with its own jet, hence the matrix element for this
reaction can be used to describe the process:
\begin{equation}
\Big(l\bar{l'} \rightarrow\Big) V \rightarrow n\;{\rm jets},
\label{revtwo}
\end{equation}
but also, after appropriate crossings and convolution with parton
distribution functions, the reactions
\begin{eqnarray}
h h' & \rightarrow & V(\rightarrow l\bar{l'})+(n-2) 
\;{\rm jets}\, , \label{revthree} \\
l h & \rightarrow &l'+(n-1) \,{\rm jets},
\end{eqnarray}
where $h$ and $h'$ are hadrons.
One may also consider reactions without the vector boson $V$ in (\ref{revthree})
and describe pure QCD reactions like:
\begin{equation}
h h' \rightarrow (n-2) \;{\rm jets}.
\label{revfive}
\end{equation}
These reactions (\ref{revthree}-\ref{revfive}) represent 
the most prominent jet cross sections
measured at colliders. When one performs the phase space integral over
the final state parton momenta numerically, it is clear that at lowest
order, where each parton forms a separate jet, one may impose any
jet-defining cuts or acceptance cuts on the calculation.
Although jet-finding algorithms can also be implemented easily,
such a calculation cannot reproduce the details of the corresponding
experimental cross section as it does not yield a prediction
for the internal structure of a jet.

It is the main purpose of the PSS method as described in
Refs.~\cite{GG92,GGK93} to maintain these features (i.e. 
the applicability of matrix elements to different
processes, the easy numerical implementation of cuts and acceptance
criteria, and the ability to change the jet algorithm without any
analytical recalculations) at next-to-leading
order.

The reasons for incorporating next-to-leading corrections
in the modelling of jet cross sections are well-known:
\begin{itemize}
\item Reducing the normalization uncertainty via decreased
sensitivity to changes in the renormalization scale and/or
factorization scale.
\item Testing the convergence of the QCD perturbation
series for the observable under investigation.
\item Modelling of inter-jet hadronic radiation.
\item Modelling sensitivity to the jet algorithm.
\end{itemize}
The latter occurs because of the presence of the $n+1$
parton matrix element in the next-to-leading corrections.
It is now possible for {\em two} partons to form
a jet.  However it is up to the particular jet algorithm
to decide whether, and how, to merge a given pair of partons into a jet.

For definiteness we shall consider the process:
\begin{equation}
V \rightarrow  q\bar{q}+(n-2)g
\label{revone2}
\end{equation}
where we assume that the vector boson has been produced in an $e^+e^-$
collision.  For full details, we refer the reader 
to Ref.~\cite{GG92}. Later in
this section we describe how to make the transition to the other
reactions (\ref{revthree}-\ref{revfive}).  That part of the discussion
is a synopsis of Ref.~\cite{GGK93}.

At next-to-leading order, the contribution of the $n+1$-parton matrix
element to the $n$-jet cross section in $e^+e^-$ collisions involves
integration over configurations where one parton is soft or two
are collinear.  In such situations divergences arise.  They
cancel against corresponding divergences in the virtual corrections 
to the $n$-parton contribution.  The PSS method isolates these 
divergences by
slicing the $n+1$-parton phase space into ``hard'' and ``soft plus
collinear'' regions.

The hard region of phase space is essentially defined such that all of the $n+1$
partons are resolved: they are separated in phase space such that the invariant mass
of any given two partons is larger than an arbitrary (small)
theoretical cutoff $\smin$, $s_{ij}=(p_i+p_j)^2=2 p_i\cdot p_j > \smin$, where 
$p_i$ and $p_j$
represent the momenta of partons $i$ and $j$.  In this region of phase space the
$n+1$-parton cross section is finite and all phase space integrals
may be done numerically.  We may write this contribution as
\begin{equation}
d\sigma^R(e^+e^-\rightarrow (n+1)\;\rpartons) = \frac{1}{2^2}\frac{1}{2\Q^2} 
\Big|M(1,\ldots,n+1)\Big|^2
dP^{R}(\Q;1,\ldots,n+1)\, ,
\label{dsR}
\end{equation}
where the spin averages and the flux factor have been 
shown explicitly, and $|M(..)|^2$ is the $n+1$-parton LO matrix element squared and 
spin-summed.
The resolved phase space in Eq.~(\ref{dsR}) is given by
\begin{eqnarray}
dP^{R}(\Q;1,\ldots,n+1)&=&\frac{1}{n_g!\Pi_f n_q^f!\Pi_f n_{\bar q}^f!}
\prod_{ij} 
\theta(s_{ij}-\smin) dP(\Q;1,\ldots,n+1)\, ,
\label{hphsp}
\end{eqnarray}
where $\Q$ stands for the sum of the momenta of the incoming electron
and positron 
and $dP$ is the usual phase space measure in 4 dimensions:
\begin{equation}
dP(\Q;1,\ldots,n+1) = 
\Big[\prod_{i=1}^{n+1}{d^{3}{\bf p}_i\over(2\pi)^{3} 2E_i}\Big]
(2\pi)^4 \delta(\Q-\sum_i^{n+1} p_i) \nonumber \\
\end{equation}
An identical particle factor for gluons and (anti)quarks
of the same flavor is included. The product over $\theta$ functions
is defined such that all non-divergent, finite contributions 
are properly included in this resolved contribution. 
See below Eq.~(\ref{eq:cross}) for more detail.

In order to understand the behavior of the matrix elements in 
the soft and collinear region it is very useful to decompose
the amplitude for producing $n+1$ partons in the final state
using color-ordering~\cite{Colorordering}. To illustrate this, let us consider
here for definiteness the amplitude for $\ng+1=n-1$
gluons radiating off the colored quark line that couples to the vector
boson $V^{\mu}$,
\beq
M(1,\ldots,n+1)_{c_1 c_2} = V^{\mu}\hat{S}_{\mu}(K;1,\ldots,\ng+1;\overline{K})_{c_1 c_2}\, ,
\eeq
where $K$, $\overline{K}$ are the momenta 
of the final state quark and antiquark respectively and $c_1, c_2$ are 
their color indices.
One may write
\begin{equation}
\hat{S}_{\mu}(K;1,\ldots,\ng+1;\overline{K})_{c_1 c_2} = 
i e g^{\ng+1} \sum_{P(1\ldots \ng+1)}(T^{a_1}\ldots T^{a_{\ng+1}})_{c_1 c_2}
 S_{\mu}(K;1,\ldots,\ng+1;\overline{K})\,.
\label{colordec}
\end{equation}
Here $S_{\mu}(K;1,\ldots,\ng+1;\overline{K})$ is a colorless
subamplitude in which the $\ng+1$ gluons couple in a color ordered way to
the quark line, and $e,g$ are the QED and QCD coupling constants
respectively.  The full amplitude is the sum over all permutations of
these gluons. 

A similar decomposition may be performed for the amplitude with four
or more quarks in the final state.  We shall not discuss such amplitudes 
in any detail.  We merely note that such
amplitudes are color-suppressed in relation to the gluonic ones, and
can be straightforwardly included in the formalism (see Ref.~\cite{GG92} for
details).

Using Eq.~(\ref{colordec}), the square of the full amplitude can be expanded in
the number of colors $N$. One finds, after contraction with the vector boson $V$,
and factoring out an overall factor $(N^2-1)/N$
\begin{equation}
\Big| V^{\mu}\hat{S}_{\mu}\Big|^2 = 
e^2\Big(\frac{g^2N}{2}\Big)^{\ng+1}\Big(\frac{N^2-1}{N}\Big)
\Big(A[\ng+1]+\frac{1}{N^2}B[\ng+1]+\ldots)\, ,
\label{squareamp}
\end{equation}
where
\beq
A[\ng+1] = \sum_{P(1\ldots \ng+1)} |V^{\mu}S_{\mu}
(K; 1\ldots \ng+1;\overline{K})|^2\ .
\eeq
There is no known general expression for the term $B[\ng+1]$, but
for any given $\ng$ it may be computed straightforwardly \cite{BG}.
Note also that Eq.~(\ref{squareamp}) is a finite series.
For example, with $\ng=1$ only the $A[2]$ and $B[2]$ terms are present,
and $B[2]$ reads
\beq
B[2] = -| V^{\mu}(\,S_{\mu}(K; 1,2;\overline{K})+S_{\mu}(K; 2,1;\overline{K})\,)|^2\, .
\label{Btwo}
\eeq 
Here the 3-gluon coupling cancels, and the gluons behave in fact
like photons \cite{BG}.    

We now examine the complementary soft plus collinear region where one
or two of the $s_{ij}$ are smaller than the theoretical cut-off,
$\smin$.  In this region only $n$ of the $n+1$ partons are resolved.
The divergences that occur in this region are regularized by working in
$d=4-2\epsilon$ dimensions.  
The advantage of the color-ordered
decomposition of the matrix element is that an $n+1$-parton colorless 
ordered subamplitude factorizes into an $n$-parton amplitude
and a $1$-parton factor, for an appropiately small value of $\smin$. 
The $n+1$-parton phase space also factorizes into the $n$-parton
phase space and a $1$-parton factor. The integration of the $1$-parton 
amplitude factor over the $1$-parton phase space
can easily be performed analytically.

After this integration the squared soft and collinear matrix elements,
resp. $| V^{\mu}\hat{S}_{\mu}|^2_S$ and $|V^{\mu}\hat{S}_{\mu}|^2_C$
(they are given explicitly in Eqs.~(\ref{eq:soft}) and (\ref{eq:coll})), 
can be combined with the leading order (LO) and NLO virtual (V)
$n$-parton contributions to form an effective $n$ resolved parton
squared matrix element:
\beq 
| M(1,\ldots,n)|^2_{\rm eff} = |
M(1,\ldots,n)|^2_{\rm LO+V} + | V^{\mu}\hat{S}_{\mu}|^2_S +
|V^{\mu}\hat{S}_{\mu}|^2_C\, .
\label{Mneff}
\eeq
After coupling constant renormalization $|M(1,\ldots,n)|^2_{\rm eff}$
is finite, by virtue of the renormalizability of QCD~\cite{thooft} and
the KLN theorem~\cite{KLN}.

The resolved $n$ parton contribution is now given by: 
\begin{equation}
d\sigma^R(e^+e^-\rightarrow n\;\rpartons) = \frac{1}{2^2}\frac{1}{2\Q^2} \Big|
M(1,\ldots,n)\Big|^2_{\rm eff}
dP^R(\Q;1,\ldots,n)\, .
\label{dsR2}
\end{equation}
Note that because $\Big| M(1,\ldots,n)\Big|^2_{\rm eff}$ is finite, we
may take the $n$-parton phase space measure here in 4 dimensions.  

Combining Eqs.~(\ref{dsR2}) and (\ref{dsR}),
we may now write the next-to-leading order $n$-jet cross section in 
$e^+e^-$ collisions as 
\begin{equation}
d\sigma(e^+e^-\rightarrow n\;\rjets) =
\Theta\Big[ d\sigma^R(e^+e^-\rightarrow n\;\rpartons) +
\int  d\sigma^R(e^+e^-\rightarrow (n+1)\;\rpartons)\Big].
\label{eq:cross}
\end{equation}
Here $\Theta$ represents all the experimental effects, including the
jet algorithm.  The integral in the second term represents the
projection of the phase space 
of the $n+1$ resolved partons onto the $n$
jets phase space. We can now be more explicit about the definition
of the product over the $\theta$-functions in Eq.~(\ref{hphsp}), which
is implicit in the second term in Eq.~(\ref{eq:cross}). 
For each color order contributing to $|M(1,\ldots,n+1)|^2$,
the product runs over no more than two invariants involving the same
momentum, otherwise some finite contributions of order $\smin/\smin$
might not be properly included.

Eq.~(\ref{eq:cross}) now involves only finite
quantities and may be used to construct a fully differential NLO Monte
Carlo program for the production of $n$ jets in $e^+ e^-$
collisions. Note that both terms on the right hand side of
Eq.~(\ref{eq:cross}) depend on $\smin$, but that no observable may
depend on it, as long as $\smin$ is taken small
enough such that the approximations used in the soft and collinear
region are valid.  For a given process this must be checked
numerically with the Monte Carlo program.

We now briefly describe the procedure to calculate the soft and collinear 
contributions.  

\subsection{Soft Behavior of Matrix Element, Phase Space, and Cross Section}

Colorless ordered subamplitudes share the property of QED amplitudes that,
in the limit where one of the external gauge bosons becomes soft, the
amplitude factorizes into a lower order amplitude and an eikonal
factor:
\beqn S_\mu (K;1,...,\ng,s;\overline{K}) &
\rightarrow &\,e_\lambda(\ng;s;\overline{K})\; S_\mu
(K;1,...,\ng;\overline{K})\, , \nonumber\\ S_\mu
(K;1,...,m,s,m+1,...,\ng;\overline{K}) & \rightarrow &\,
e_\lambda(m;s;m+1)\; S_\mu (K;1,...,\ng;\overline{K})\, , \nonumber \\
S_\mu (K;s,1,...,\ng;\overline{K}) & \rightarrow &\, e_\lambda(K;s;1) \;
S_\mu (K;1,...,\ng;\overline{K})\, , 
\eeqn where the eikonal factor is
given by 
\beq e_\lambda(a;s;b)=\left(
\frac{\eps_\lambda(s)\cdot p_a}{p_a\cdot p_s}-\frac{\eps_\lambda(s)\cdot p_b}
{p_b\cdot p_s} \right).  
\eeq
Here $\eps_\lambda^\mu(s)$ is the polarization vector of the soft
gluon with momentum $s$ and helicity $\lambda$, and the momenta $p_a$
and $p_b$ are those of the two hard partons adjacent to the soft gluon in
the specific color order considered.  Summing over 
the helicities of the soft gluon
one obtains for the square of the amplitude
\begin{eqnarray}
| V^{\mu}\hat{S}_{\mu}|^2 &\rightarrow& 
e^2\Big(\frac{g^2N}{2}\Big)^{\ng}\Big(\frac{N^2-1}{N}\Big)
\Big[\sum_{P(1\ldots \ng)} s_F(K;1,\ldots,\ng;\overline{K})
\,| V^{\mu}S_{\mu}(K;1,\ldots,\ng;\overline{K})|^2 \nonumber \\
&+&O(\frac{1}{N^2})\Big] \, ,
\label{softphme}
\end{eqnarray}
where 
\beq
s_F(K;1,\ldots,\ng;\overline{K}) = \Big(\frac{g^2N}{2}\Big)\Big[
\sum_{\lambda}|e_\lambda(K;s;1)|^2+\ldots+ \sum_{\lambda}|e_\lambda(\ng;s;\overline{K})|^2\Big]\, .
\label{littles}
\eeq
In the soft limit the $d$-dimensional ($d=4-2\epsilon$) 
$n+1$-parton phase space measure,
\begin{equation}
dP^d(\Q;1,\ldots,n+1) = 
\Big[\prod_{i=1}^{n+1}{d^{d-1}{\bf p}_i\over (2\pi)^{d-1} 2E_i}\Big]
(2\pi)^d \delta(\Q-\sum_i p_i)\,,\\
\end{equation} 
also factorizes, up to terms of order $\smin/\Q^2$, according to
\beq
dP^d(\Q;1,\ldots,a,s,b,\dots,n+1) \rightarrow dP^d(\Q;1,\ldots,n)\ 
dP^{\eps}_{soft}(a,s,b)\, ,
\label{softphspfact}
\eeq
where 
\begin{eqnarray}
dP^{\eps}_{soft}(a,s,b) &=& \frac{(4\pi)^{\epsilon}}{16\pi^2\Gamma(1-\epsilon)}
\frac{ds_{as}ds_{bs}}{s_{ab}}\Big[\frac{s_{as}s_{bs}}{s_{ab}}\Big]^{-\epsilon}
\theta(\smin-s_{as})\theta(\smin-s_{bs})\ .
\label{softps}
\end{eqnarray}
The $\theta$-functions define the soft region formally.  
After combining (\ref{softphme}), (\ref{littles}) and (\ref{softps}), and 
integrating over $s_{as}$ and $s_{bs}$ the effective squared matrix element
with $\ng$ resolved gluons and two resolved quarks is
\begin{eqnarray}\label{eq:soft}
| V^{\mu}\hat{S}_{\mu}|^2_S &=& e^2\Big(\frac{g^2N}{2}\Big)^{\ng}
\Big(\frac{N^2-1}{N}\Big)
\Big[ \sum_{P(1\ldots \ng)} S_F(K;1,\ldots,\ng;\overline{K})
| V^{\mu}S_{\mu}(K;1,\ldots,\ng;\overline{K})|^2 \nonumber \\
&+&O(\frac{1}{N^2}) \Big] \, ,
\end{eqnarray}
with 
\beq
S_F(K;1,\ldots,\ng;\overline{K}) = \Big(\frac{\alpha_s N}{2 \pi}\Big)\frac{1}{\Gamma(1-\epsilon)}
\Big(\frac{4\pi\mu^2}{\smin}\Big)^\eps\frac{1}{\epsilon^2}
\Big[
\Big(\frac{s_{K1}}{\smin}\Big)^{\epsilon}+\ldots+ 
\Big(\frac{s_{\ng\overline{K}}}{\smin}\Big)^{\epsilon}\Big]\, .
\label{softfac}
\eeq Here $\mu$ is the dimension regularization scale, and $\alpha_s =
\alpha_s(\mu)=g^2\mu^{-2\epsilon}/4\pi$.  We may interpret the
dimensional regularization scale $\mu$ here already as the
renormalization scale $\mu_R$, as coupling constant renormalization is
implicitly assumed.  Note that the identical particle factor in
Eq.~(\ref{hphsp}), $1/(n_g+1)!$, is changed to $1/n_g!$ 
in Eq.~(\ref{softphspfact}), thereby taking into account the $(n_g+1)$
possibilities of choosing a soft gluon.  The contribution to 
$| V^{\mu}\hat{S}_{\mu}|^2_S$ at higher order in $1/N$ 
can be calculated straightforwardly.  

\subsection{Collinear Behavior of Matrix Element, Phase Space and Cross Section}

The matrix elements also develop a singularity when two of the partons,
say $a$ and $b$, become collinear, i.e.
\beqn
p_c=p_a + p_b \; ; \; p_a=z p_c \; ; \; p_b=(1-z) p_c \nonumber \\
s_{ab}=p_c^2=2 p_a.p_b < \smin.
\label{colmom}
\eeqn
where $\smin$ is again assumed to be small.
In this case the full squared matrix element factorizes according 
to~\footnote{In Ref.~\cite{GG92} the notation $\hat{c}_F^{ab \go c}$ was
used to indicate the splitting of parton $c$, instead of $\hat{c}_F^{c\go ab}$.}
\beq
| V^{\mu}\hat{S}_{\mu}(1,\ldots,a,b,\ldots,n+1)|^2 \go
\hat{c}_F^{c\go ab} | V^{\mu}\hat{S}_{\mu}(1,\ldots,c,\ldots,n)|^2\ .
\label{eq:mefact}
\eeq
It is useful to extract color factors via
\beq
\hat{c}_F^{c\go ab} = \Big(\frac{g^2N}{2}\Big) f^{c\go ab} \, .
\label{collfac}
\eeq
Then we have
\beq
f^{c\go ab} = \frac{P^{\eps}_{c\go ab}(z)}{s_{ab}}\, .
\label{fdef}
\eeq
The splitting functions $P^\eps_{c\go ab}$ read, in the conventional scheme (in which both
external and internal particles are treated in $d=4-2\epsilon$ dimensions)
\begin{eqnarray}
P^\eps_{g\go gg}(z) &=& 2\Bigg(\frac{1+z^4+(1-z)^4}{z(1-z)}\Bigg)\, , \nonumber\\
P^\eps_{g\go q\bar{q}}(z)&=& \frac{2}{N}\Bigg(\frac{z^2+(1-z)^2-\epsilon}{1-\epsilon}\Bigg)\, , \nonumber\\
P^\eps_{q\go qg}(z) &=& 2(1-\frac{1}{N^2})\Bigg(\frac{1+ z^2-\epsilon(1-z)^2}{1-z}\Bigg) \, .
\label{eq:Peps}
\end{eqnarray}
Notice that only when two adjacent partons in the subamplitude in 
a specific color order 
become collinear does one find a divergence.
Here, the four quark ($+ (n_g-1)$ gluons) amplitude,  $\hat{T}_\mu$, must be considered:  
when a neighboring quark and antiquark form a flavor singlet and become 
collinear, then this 
amplitude also contributes to the $n$ parton amplitude.  
In analogy to Eq.~(\ref{softphme}), we have therefore in the collinear limit: 
\begin{eqnarray}
\Big| V^{\mu}\hat{S}_{\mu}\Big|^2 &+& \Big| V^{\mu}\hat{T}_{\mu}\Big|^2 \go
e^2\Big(\frac{g^2N}{2}\Big)^{\ng}\Big(\frac{N^2-1}{N}\Big)  \nonumber \\
 &\times& \Big[ \sum_{P(1\ldots \ng)} c_F(K;1,\ldots,\ng;\overline{K})
\,| V^{\mu}S_{\mu}(K;1,\ldots,\ng;\overline{K})|^2  
+ O(\frac{1}{N^2}) \Big]\, ,
\end{eqnarray}
where
\beq
c_F(K;1,\ldots,\ng;\overline{K}) = \Big(\frac{g^2N}{2}\Big)\Big[
f^{K\go qg}+f^{1\go gg}+\ldots+ f^{\overline{K}\go gq}
+\ng\, n_f f^{g\go q\bar{q}}\Big]\,.
\label{collme}
\eeq
The last term in this expression is due to $\hat{T}_\mu$.  
The factor $\ng$ in front of this last term arises because the identical
particle factor in Eq.~(\ref{hphsp}) for the four quarks contribution
changes from $1/(\ng-1)!$ to $1/\ng!$.  The factor $n_f$ results from
summing over all (light) quark flavors.  
In the collinear region, defined in Eq.~(\ref{colmom}), the phase space measure also factorizes, 
up to terms of order $\smin/\Q^2$ as:
\beq
dP^d(\Q;1,\ldots,n-1,a,b) \rightarrow dP^d(\Q;1,\ldots,n-1,c)\ 
dP^{\eps}_{coll}(a,b;z)\ , 
\label{colphspfact}
\eeq
where
\beq
dP^{\eps}_{coll}(a,b;z) = \frac{(4\pi)^{\epsilon}}{16\pi^2\Gamma(1-\epsilon)}
ds_{ab}dz\Big[s_{ab}z(1-z)\Big]^{-\epsilon}
\theta(\smin-s_{ab})\,.
\label{collphsp}
\eeq
To obtain the collinear behavior of the cross section one needs to integrate 
the following expression:
\begin{eqnarray}
\int  \Big(\frac{g^2N}{2}\Big) f^{c\go ab}  dP^{\eps}_{coll}(a,b;z)
&\equiv&  \Big(\frac{\alpha_s N}{2 \pi}\Big)\frac{(4\pi\mu^2)^{\epsilon}}{\Gamma(1-\epsilon)} \nonumber \\
&\times&  \int_0^{\smin} ds_{ab}\, s_{ab}^{-1-\epsilon}
\Big[\frac{1}{4}\int_{z_1}^{1-z_2}dz \Bigg[z(1-z)\Big]^{-\epsilon}P^\eps_{c\go ab}(z)\Bigg]
\nonumber \\ &=&  - \Big(\frac{\alpha_s N}{2 \pi}\Big)\frac{1}{\Gamma(1-\epsilon)}
\Big(\frac{4\pi\mu^2}{\smin}\Big)^\eps\frac{1}{\epsilon}\, I_{c\go ab}(z_1,z_2)\, ,
\label{Idef}
\end{eqnarray}
where the boundary values $z_1$ and $z_2$ prevent the integral from picking up
contributions from the soft region:
\beq
s_{a-1\,a} > z_1\,s_{a-1\,c} = \smin \;\;,\;\;
s_{b\,b+1} > (1-z_2)\,s_{c\,b+1} = \smin \,.
\label{zlimits}
\eeq
These boundaries clearly depend on the color order of the particular subamplitude
$S_\mu$ under consideration.
The collinear behavior of the effective squared matrix element 
with $n$ resolved partons, of which
one is an unresolved collinear pair, is now given by
\begin{eqnarray}
\Big| V^{\mu}\hat{S}_{\mu}\Big|^2_C &=& 
e^2\Big(\frac{g^2N}{2}\Big)^{\ng}\Big(\frac{N^2-1}{N}\Big) \nonumber \\
&\times& \Big[ \sum_{P(1\ldots \ng)} C_F(K;1,\ldots,\ng;\overline{K})
\, | V^{\mu}S_{\mu}(K;1,\ldots,\ng;\overline{K})|^2 
+O(\frac{1}{N^2})\Big] \,
,
\label{eq:coll}
\end{eqnarray}
where 
\beqn
\label{eq:cf}
C_F(K;1,\ldots,\ng;\overline{K}) &=& -\Big(\frac{\alpha_s N}{2 \pi}\Big)\frac{1}{\Gamma(1-\epsilon)}
\Big(\frac{4\pi\mu^2}{\smin}\Big)^\eps\frac{1}{\epsilon}\Big[ 
I_{K\go qg}(0,\frac{\smin}{s_{K1}})
 \\ &+&I_{1\go gg}(\frac{\smin}{s_{K1}},\frac{\smin}{s_{12}})
+\ldots 
+ I_{\overline{K}\go g\bar{q}}(\frac{\smin}{s_{\ng\overline{K}}},0) + 
\ng\, n_f I_{g\go q\bar{q}}(0,0)\, .
\Big] \nonumber
\eeqn
Note that when the boundary corresponds to a soft fermion,
there is no singularity, so we can put the corresponding $z_i$ to zero.
For completeness we list here the results for the $I$ functions,
defined in Eq.~(\ref{Idef}),
for massless partons \cite{GG92,GGK93}.
\beq
I_{g\go gg}(z_1,z_2) = \Big(\frac{z_1^{-\eps}+z_2^{-\eps}-2}{\eps}\Big) 
   -\frac{11}{6} + \big(\frac{\pi^2}{3}-\frac{67}{18}\Big)\eps + O(\eps^2)\, ,
\label{ifuncggg}
\eeq
\beq
I_{q\go qg}(z_1,z_2) = (1-\frac{1}{N^2})\Big[\Big(\frac{z_2^{-\eps}-1}{\eps}\Big) 
   -\frac{3}{4} + \big(\frac{\pi^2}{6}-\frac{7}{4}\Big)\eps\Big] + O(\eps^2)\, ,
\label{ifuncqqg}
\eeq
\beq
I_{g\go q\bar{q}}(z_1,z_2) = \frac{1}{N}\Big[\frac{1}{3} + \frac{5\eps}{9}\Big] + O(\eps^2)\,.
\label{ifuncgqq}
\eeq
For $I_{q\go qg}(z_1,z_2)$ the $1/N^2$ term in fact contributes to the
${\cal O}(1/N^2)$ terms in Eq.~(\ref{eq:coll}).  We use this 
definition of $I_{q\go qg}$
because it is more convenient for what follows.
The remaining $I$'s are trivially derived from these, via the relations
\beq
I_{q\go gq}(z_1,z_2) = I_{q\go qg}(z_2,z_1) \;\;,\;\;
I_{g\go \bar{q}q}(z_1,z_2) = I_{g\go q\bar{q}}(z_2,z_1)
\label{ifuncder1}
\eeq
and 
\beq
I_{\bar{q}\go \bar{q}g}(z_1,z_2) = I_{q\go qg}(z_1,z_2) \;\;,\;\;
I_{\bar{q}\go g\bar{q}}(z_1,z_2) = I_{q\go gq}(z_1,z_2) \, .
\label{ifuncder2}
\eeq
Again, we refer to Ref.~\cite{GG92} for all details.

\subsection{Crossing}

In this subsection we review how to describe reactions
involving initial state hadrons with the PSS method.  The original
derivation and all details can be found in Ref.~\cite{GGK93}.

The calculation of the NLO matrix elements in the previous section was
done with all partons in the final state.  To describe processes with
initial state partons at NLO, one would like to generalize the
crossing property of the LO matrix elements to NLO.  In particular,
one would like to use the already calculated effective
matrix elements of Eq.~(\ref{Mneff}), and do the crossing numerically,
without having to redo the analytical calculation for the specific process
under consideration.  
This is possible, via universal ``crossing functions'' \cite{GGK93}.
The feasibility of thus extending crossing to NLO rests again upon the
simultaneous factorization of phase space and matrix element for both
initial and final state collinear radiation.

In general the NLO differential cross section for a process with initial
hadrons $\H_1$ and $\H_2$ may be written as
\beq
d\sigma_{\H_1\,\H_2} = \sum_{a,b}\int dx_1\int dx_2
\, {\cal F}_a^{\H_1}(x_1) \,{\cal F}_b^{\H_2}(x_2) d\sigma_{ab}^{NLO}(x_1,x_2)\ ,
\label{factone}
\eeq
where $a,b$ denote parton flavors and $x_1,x_2$ are parton momentum fractions.
The symbols in this equation do not have quite the same meaning as they do in 
the standard factorization formula \cite{CSS}. Here ${\cal F}_a^{\H}(x)$ is an 
``effective'' parton distribution function and $d\sigma_{ab}^{NLO}$ is 
{\it not} the next-to-leading partonic cross section
(which we denote by $\widehat{d\sigma_{ab}^{NLO}}$). Rather,
it consists of the finite effective all-partons-in-the-final-state matrix elements,
in which partons $a$ and $b$ have simply been crossed to the initial state,
i.e. in which their momenta $p_a$ and $p_b$ have been replaced by 
$-p_a$ and $-p_b$.
The contributions that distinguish $d\sigma_{ab}^{NLO}$
from $\widehat{d\sigma_{ab}^{NLO}}$ are in fact
included in ${\cal F}_a^{\H}(x)$ and are what distinguish 
the latter from the usual
parton distribution functions $f_a^{\H}(x)$.
Explicitly, we have 
\beq
d\sigma_{ab}^{NLO} = d\sigma_{ab}^{LO} +
\alpha_s d\Big(\delta \sigma_{ab}^{NLO}\Big) + O(\alpha_s^2), 
\eeq
with $\alpha_s = \alpha_s(\mu_R)$ from the crossed matrix
elements, and
\beq
{\cal F}_a^{\H}(x) = f_a^{\H}(x,\mu_F) + \alpha_s C_a^{\H}(x,\mu_F)
 + O(\alpha_s^2),
\eeq
where mass factorization has taken place at scale $\mu_F$.
$f_a^{\H}(x,\mu_F)$ is a NLO parton distribution function 
and $C_a^{\H}(x,\mu_F)$ is a finite, 
process-independent crossing function.  

There are two contributions
that distinguish $d\sigma_{ab}^{NLO}$ from  
the full next-to-leading partonic cross section $\widehat{d\sigma_{ab}^{NLO}}$. 
First, when crossing a parton into the initial state we 
must correct for the possibility that this parton is in fact an unresolved collinear pair.
Such ``fusing'' processes are not to be taken into account at leading twist
and must hence be subtracted. 
Second, the contribution from the initial state collinear region  must be included.  
Schematically, one has therefore, before mass factorization,
\beq
C_a^{\H}(x) \sim \sum_{c,b}\Big[ \int_x^1\frac{dz}{z} 
f_c^{\H}(\frac{x}{z}) P_{c\go ab}(z)
- f_a^{\H}(x)\int_0^1dz P_{a\go cb}(z)\Big]\frac{\smin^{-\eps}}{\eps}\,,
\eeq
where the $f_a^{\H}$ are the bare distribution functions and 
the collinear singularities are still present.
After mass factorization in a particular scheme, e.g. $\overline{\rm MS}$, one may write
\beq
C_a^{\H,\msb}(x,\mu_F) = \frac{N}{2\pi}\Big[A_a^\H(x,\mu_F)\ln(\frac{\smin}{\mu_F^2})
+B_a^{\H,\msb}(x,\mu_F)\Big]\, .
\label{cdefab}
\eeq
The functions $A$ and $B$ can be expressed as convolution integrals
over the parton distribution functions and are listed in
Ref.~\cite{GGK93}.  The $B$ functions are scheme dependent.  The
crossing functions, being process independent (universal), need to be calculated only once
for any set of parton distribution 
functions (see \cite{GRV,CTEQ,MRS} for recent sets).

The full NLO differential cross section can now be written as:
\begin{eqnarray} \label{eq:dshh}
d\sigma_{\H_1\,\H_2} &= &\sum_{a,b}\int dx_1\int dx_2
 f_a^{\H_1}(x_1,\mu_F) \,f_b^{\H_2}(x_2,\mu_F) 
  \, d\sigma_{ab}^{NLO}(x_1,x_2) + \nonumber \\
\hspace*{-5cm}&+& 
\alpha_s (\mu_F)
\Big( C^{\H_1}_a(x_1,\mu_F) f_b^{\H_2}(x_2,\mu_F) 
+ f_a^{\H_1}(x_1,\mu_F) C^{\H_2}_b(x_2,\mu_F)
\Big) d\sigma_{ab}^{LO}(x_1,x_2) \nonumber \\
&=& \sum_{a,b} 
\int dx_1\int dx_2
 f_a^{\H_1}(x_1,\mu_F) \,f_b^{\H_2}(x_2,\mu_F) \,\widehat{d\sigma_{ab}^{NLO}}
\,
\end{eqnarray}
where the scheme dependence of the individual functions is suppressed.

We have reviewed quite extensively the PSS method for the calculation
of general jet cross sections involving massless partons, and now proceed
to the main task of this paper, which is to generalize the method to
include fragmentation functions (in the next section), and heavy
quarks (in section 4).

\section{Fragmentation and Tagging Functions}

The PSS method can be extended to describe not just jet cross
sections, but also cross sections in which a particular particle is
identified or ``tagged" in the final state.  Here that particle is
assumed to be a hadron $\H$.  Descriptions for such processes require
the introduction of a fragmentation function $D_h^\H(z)$, representing
the probability of a parton $h$ to fragment into $\H$, with $z$ the
momentum fraction of $h$ carried by $\H$.  In this section we show how
to include fragmentation functions in the PSS method.  

Let us consider the NLO cross section for the production of a hadron $\H$ 
and $(n-1)$ jets in $e^+ e^-$ collisions.  

\beq
d\sigma_\H=\sum_h d\sigma_h^{NLO} \cdh (z) dz\, ,
\label{eq:sh}
\eeq
where \cdh\ is an {\it effective} next-to-leading order fragmentation function,
in a similar sense that ${\cal F}_a^{\H}$ in Eq.~(\ref{factone}) was an
effective next-to-leading order parton distribution function.
$d\sigma_h^{NLO}$ is in fact given by Eq.~(\ref{eq:cross}), with one of the resolved
partons fixed to have flavor $h$. 
The sum is over all parton flavors $h$ in the intermediate
 state.   The goal is again to use the
results already obtained for the production of $n$ jets at NLO 
without need to modify
the effective matrix element for the resolved $n$-parton contribution
(Eq.~(\ref{Mneff})).  
We may expand
\beq
d\sigma_h^{NLO}=d\sigma_h^{LO} +\als d\Big(\delta \sigma_h^{NLO}\Big) + O(\als^2)\, ,
\label{eq:shn}
\eeq
\beq
\cdh (z)=\dh (z)+\als \ch (z) + O(\als^2)\, ,
\label{eq:dh}
\eeq
where the \ch\ are the final state equivalents of the
crossing functions, described in the previous section.  We 
call them ``tagging functions''.  

Substitution of Eqs.~(\ref{eq:shn}) and (\ref{eq:dh}) into (\ref{eq:sh}) gives
up to order $\alpha_s$
\beq
d\sigma_\H= \sum_{h} \Big[(d\sigma_h^{LO} +\als 
d\Big(\delta \sigma_h^{NLO}\Big)) \dh (z) 
 +\als d\sigma_h^{LO} \ch (z)\Big]dz\,.
\label{eq:sh2}
\eeq
What follows is in fact very similar to the derivation of the crossing
functions in Ref.~\cite{GGK93}. 
The tagging functions consist of those corrections 
that render Eq.~(\ref{eq:sh2}) correct.  

In the evaluation of the contribution in the collinear region
in section 2, the integration over the momentum fraction 
$z$  was performed.  This collinear integration should however have been
done in convolution with the fragmentation functions.
Therefore, the {\em integrated} contribution must first be subtracted,
and then the properly convoluted contributions should be added.
For a specific color ordering, the integrated contribution is given by:
\beq
\als T_{h,int}^\H(z) = 
\sum_{p,u}\int \hat{c}_F^{h\go pu}(y) dP^\eps_{coll}(p,u;y) 
 D_h^\H(z)\, ,
\label{eq:basictagint}
\eeq
so that
\beq
T_{h,int}^\H(z)=- \fac \sum_{p,u} I_{h\go pu}(y_1,y_2) D_h^\H(z),
\label{eq:cfhint}
\eeq
where the sum is over all allowed parton flavors.
The $I_{h\go pu}$ are given in
Eqs.~(\ref{ifuncggg}-\ref{ifuncder2}).  Here $y_1$ and $y_2$ are the
equivalent of the color-order dependent boundaries $z_1$ and $z_2$ in
Eq.~(\ref{zlimits}).  The complete tagging
functions $T_{h}^\H$ will be independent of the color ordering.

Next, we need to add the collinear terms having the proper convolution with 
fragmentation functions.  For each specific color ordering, 
the resulting coefficient  is
\beq
\als T_{h,conv}^\H(z) dz= \sum_{p,u}\int dP^\eps_{coll}(p,u;y) \hat{c}_F^{h\go pu}(y)
 D_p^\H(x) \delta(z-xy) dx\, ,
\label{eq:basictagconv}
\eeq
where $y=E_p/E_h$ and $x=E_\H/E_p$, so that $z=E_H/E_h$ is an energy
fraction. Using the expressions of section 2, we find
\beq
T_{h,conv}^\H (z) =-\fac  \sum_{p,u} \int_{z}^{1-y_2} {dy \over y}
\Big[y(1-y)\Big]^{-\eps} 
\frac{1}{4}\,  P^\eps_{h\go pu}(y) D_p^\H(z/y) \, .
\eeq
Other definitions of $z$ are equivalent, e.g. using the plus components rather than the energies.
We have
\beq
T_{h,conv}^\H (z)=-\fac \sum_{p,u} \int_{z}^{1-y_2} \frac{dy}{y} \Big[y(1-y)\Big]^{-\eps}
          \frac{1}{4}\,  P^\eps_{h\go pu}(y) D_p^\H(z/y)\, .
\label{ffunint}
\eeq
This may be written as
\beq
T_{h,conv}^\H (z)=-\fac \sum_p \int_{z}^{1} \frac{dy}{y} K_{h\go p}(y,y_2) 
                D_p^\H(z/y)\, ,
\label{Tconv}
\eeq
with
\beq
K_{h\go p}(y,y_2) = \frac{1}{4} y^{-\eps} (1-y)^{-\eps}  P^\eps_{h\go pu}(y)
\theta(1-y-y_2)\, .
\eeq
The functions $K_{h\go p}(y,y_2)$ are equal to the functions $J_{p\go h}(y,y_2)$ 
in Ref.~\cite{GGK93} in the crossing case, but multiplied by $y^{-\eps}$
and  with $h$ and $p$ inverted.
Using the plus prescripton
\beq
F(z)_+ \equiv \lim_{\beta\rightarrow 0}\Big(
\theta(1-z-\beta) F(z) - \delta(1-z-\beta)\int_0^{1-\beta} dy F(y)\Big)\, ,
\label{plusdef}
\eeq
we may expand the $K_{h\rightarrow p}(y,y_2)$ in order to extend
the $y$-integral to 1. For example, the right hand side in Eq.~(\ref{Tconv})
contains terms that may be expanded as
\beqn
\int_z^{1-y_2} dy \frac{g(y)}{(1-y)^{1+\eps}}
&=& \Big(\frac{y_2^{-\eps}-1}{\eps}\Big) g(1) 
+ \int_z^1 dy \frac{g(y)}{(1-y)_+} \\
\hspace{-4cm} &-& \epsilon \int_z^1 dy\, g(y)\Bigg(\frac{\ln(1-y)}{1-y}\Bigg)_+
+ {\cal O}(\eps^2) \nonumber \, .
\eeqn

Now combining the integrated and convoluted components leads to
\beqn
T_{h}^\H (z)&=&T_{h,conv}^\H (z)-T_{h,int}^\H(z) \nonumber\\
&\equiv&\sum_p \int_{z}^{1} \frac{dy}{y} Y_{h \go p}(y) D^\H_p(z/y)\, .
\label{Tfinal}
\eeqn
The $Y_{h \go p}(y)$ are independent of the color ordering-dependent
boundaries $y_1$ and $y_2$, so that $T_{h}^\H$ is indeed an overall
factor multiplying the LO partonic cross section, as advertised in
Eq.~(\ref{eq:sh2}).

The functions $T_{h}^\H (z)$ in (\ref{Tfinal}) still contain collinear
singularities.  These singularities are cancelled via mass
factorization applied to the fragmentation functions, which in the present case amounts to writing 
\beq
D_h^\H(z)=D_h^{\H,{\rm scheme}}(z,\mu_D)+ \als \sum_p \int_z^1
\frac{dy}{y} R_{h\go p}^{\rm scheme}(y,\mu_D)D_p^\H(z/y,\mu_D)\,
+O(\als^2)\,, \eeq 
where $\mu_D$ is the final state factorization
scale.  The $R_{h\go p}(y,\mu_D)$ are counterterm functions, the finite
terms of which determine the factorization scheme.  In this section,
where we are only dealing with massless partons, we choose the
$\overline{\rm MS}$ counterterm functions $R_{h\go p}$, as in
Ref.~\cite{GGK93} for the crossing functions: 
\beq
R^{\msb}_{q\go q}(y,\mu_D) = \facml 
\frac{1}{2}(1-\frac{1}{N^2}) \Bigg[\Big(\frac{1+y^2}{1-y}\Big)\Bigg]_+ 
\,,
\label{roneml}
\eeq
\beq
R^{\msb}_{q\go g}(y,\mu_D) = \facml \frac{1}{4} P^{\eps=0}_{q\go gq}(y) \,,
\label{rtwoml}
\eeq
\beq
R^{\msb}_{g\go q} (y,\mu_D)= \facml \frac{1}{4} P^{\eps=0}_{g\go q\bar{q}}(y)\,,
\label{rthreeml}
\eeq 
\beqn
R^{\msb}_{g\go g}(y,\mu_D) &=& \facml \Bigg\{
{(11 N -2n_f)\over 6N}\delta(1-y) \nonumber \\
&+& 2\left( {z\over(1-z)_+}+{(1-z)\over z} + z(1-z)\right)\Bigg\} \,.
\label{rfourml}
\eeqn
where $P^{\eps=0}_{c\go ab}$ can be inferred from the massless splitting functions
in Eq.~(\ref{eq:Peps}).  The counterterm functions in 
Eqs.~(\ref{roneml}-\ref{rfourml}) cancel the
$1/\epsilon$ poles and the factor
$(4\pi)^{\epsilon}/\Gamma(1-\epsilon)$, present in $Y_{h \go p}$, and
replace the dimensional regularization scale $\mu$ by the final state
factorization scale $\mu_D$.  Then we have \beq T_{h}^{\H,\overline{\rm
MS}} (z,\mu_D)=\sum_p \int_z^1 \frac{dy}{y} \Bigg(Y_{h\go
p}(y)+R^{\overline{\rm MS}}_{h\go p}(y,\mu_D)\Bigg) D_p^\H(z/y,\mu_D)
\label{taggconstr}
\eeq
Notice that the effective fragmentation function ${\cal D}$ is 
not dependent on the factorization scheme or scale $\mu_D$:
\beq
\cdh (z)=\dh (z,\mu_D)+\als T_{h}^\H (z,\mu_D) + O(\als^2)\, .
\eeq
The results of the calculation in Eq.~(\ref{taggconstr}) are summarized by
\beq
T_{h}^{\H,\overline{\rm MS}} (z,\mu_D)=\frac{N}{2\pi} 
\Bigg[U_{h}^\H(z,\mu_D) \ln(\frac{\smin}{\mu_D^2})+V_{h}^{\H,\overline{\rm MS}} (z,\mu_D)\Bigg]\, ,
\label{eq:UVdef}
\eeq
where
\beq
U_{h}^\H(z,\mu_D)=\sum_p U_{h\go p}^\H (z,\mu_D)\;\;,\;\;
V_{h}^{\H,\overline{\rm MS}} (z,\mu_D)=\sum_p V_{h\go p}^{\H,\overline{\rm MS}} (z,\mu_D) \,.
\eeq
The functions $U_{h \go p}^\H$ are scheme independent and are given by:
\beqn
U_{g \go g}^\H(z,\mu_D)&=&\int_z^1 \frac{dy}{y} 
\Bigg[\frac{(11N-2n_f)}{6N} \delta (1-y) +2 \Big(\frac{y}{(1-y)_+}+\frac{(1-y)}{y}
\nonumber\\ && +y(1-y)\Big)\Bigg] D_g^\H(z/y,\mu_D)\,,
\label{ufirst}
\eeqn
\beq
U_{q \go q}^\H(z,\mu_D)=\frac{1}{2}(1-\frac{1}{N^2})\int_z^1 \frac{dy}{y} 
 \Bigg[\frac{3}{2} \delta (1-y) +
\frac{1+y^2}{(1-y)_+}\Bigg] D_q^\H(z/y,\mu_D)\,,
\eeq
\beq
U_{g \go q}^\H(z,\mu_D)=\frac{1}{4} \int_z^1 \frac{dy}{y} 
P_{g\go q\bar{q}}^{\epsilon=0} (y) D_q^\H(z/y,\mu_D)\,,
\eeq
\beq
U_{q \go g}^\H(z,\mu_D)=\frac{1}{4} \int_z^1 \frac{dy}{y} 
P_{q\go gq}^{\epsilon=0} (y) D_g^\H(z/y,\mu_D)\,,
\eeq
where the functions $P^{\epsilon=0}_{h\go pu}$ are obtained from
Eq.~(\ref{eq:Peps}).
The scheme-dependent functions $V_{h \go p}^{\H,\overline{\rm MS}}$ can be similarly derived:
\beqn
V_{g \go g}^{\H,\overline{\rm MS}}(z,\mu_D)=\int_z^1 \frac{dy}{y} 
\Bigg[ 2 \ln y \Bigg(\frac{y}{(1-y)_+}+\frac{(1-y)}{y}+y(1-y)\Bigg) 
\nonumber \\
+\Bigg(\frac{\pi^2}{3} -\frac{67}{18} +\frac{5n_f}{9N}\Bigg) \delta (1-y) 
+2y\Big(\frac{\ln(1-y)}{1-y}\Big)_+ 
\nonumber \\
+2\Bigg(\frac{(1-y)}{y}+y(1-y)\Bigg)\ln(1-y)
\Bigg] D_g^\H(z/y,\mu_D)\,,
\eeqn
\beqn
V_{q \go q}^{\H,\overline{\rm MS}}(z,\mu_D)=\frac{1}{2} (1-\frac{1}{N^2})
\int_z^1 \frac{dy}{y} 
\Bigg[\ln y \frac{1+y^2}{(1-y)_+}
+(\frac{\pi^2}{3} -\frac{7}{2}) \delta (1-y)
\nonumber \\
+(1-y) +(1+y^2) \Bigg(\frac{\ln(1-y)}{1-y}\Bigg)_+
\Bigg] D_q^\H(z/y,\mu_D)\,,
\eeqn
\beq
V_{g \go q}^{\H,\overline{\rm MS}}(z,\mu_D)=\frac{1}{4} \int_z^1 \frac{dy}{y} 
\Bigg[P_{g\go q\bar{q}}^{\epsilon=0} (y) \ln\Big(y(1-y)\Big)
-\hat{P}_{g\go q\bar{q}}^{\eps} (y) 
\Bigg]\,D_q^\H(z/y,\mu_D)\,,
\eeq
\beq
V_{q \go g}^{\H,\overline{\rm MS}}(z,\mu_D)=\frac{1}{4} \int_z^1 \frac{dy}{y} 
\Bigg[P_{q\go gq}^{\epsilon=0} (y) \ln\Big(y(1-y)\Big)
-\hat{P}_{q\go gq}^{\eps} (y) 
\Bigg]\,D_g^\H(z/y,\mu_D)\,.
\label{vlast}
\eeq
The functions $\hat{P}_{i\go jk}^{\eps}$ are defined by 
\begin{equation}
  \label{eq:pepsdef}
  P_{i\go jk}^{\epsilon}=  P_{i\go jk}^{\epsilon=0}+\eps\,\hat{P}_{i\go jk}^{\eps}
+O(\eps^2)\,,
\end{equation}
with the $P_{i\go jk}^{\epsilon}$ listed in Eq.~(\ref{eq:Peps}).  
Notice the $\ln y$ terms generated by the extra
factor $y^{-\eps}$ in the collinear phase space measure. 
We stress that the tagging functions are universal and need
only to be calculated once for a given set of fragmentation functions
(see \cite{BKK} for recent sets).

We can combine the crossing functions and tagging functions
in order to describe a NLO fully differential cross section
for collisions of hadrons $\H_1$ and $\H_2$ in which hadron $\H$ is
tagged in the final state:
\beq
d\sigma_{\H_1\H_2\rightarrow \H+X}
=\sum_{a,b,h} \int dx_1 dx_2 dz\,{\cal F}_a^{\H1}(x_1) {\cal F}_b^{\H2}(x_2) 
                d\sigma_{abh}^{NLO}(x_1,x_2,z)\cdh (z) \,.
\label{facttwo}
\eeq
Substituting
\beq
d\sigma_{abh}^{NLO}(x_1,x_2,z)=d\sigma_{abh}^{LO} (x_1,x_2,z)+\als d\delta \sigma_{abh}^{NLO} (x_1,x_2,z)
                + O(\als^2)\,,
\eeq
\beq
{\cal F}_a^\H(x)=f_a^\H(x,\mu_F)+\als C_{a}^\H(x,\mu_F) +O(\als^2)\,,
\eeq
\beq
\cdh (z)=\dh (z,\mu_D)+\als \ch (z,\mu_D) + O(\als^2)\,,
\eeq
leads to
\beqn
d\sigma_{\H_1\H_2\rightarrow \H+X}&=& \sum_{abh} \Bigg[ 
f_a^{\H1}(x_1) f_b^{\H2}(x_2) 
\left(d\sigma_{abh}^{LO} (x_1,x_2,z)+\als d\delta \sigma_{abh}^{NLO}(x_1,x_2,z)\right) \dh (z) 
\nonumber \\
&&+\als 
C_{a}^{\H1}(x_1) f_b^{\H2}(x_2) d\sigma_{abh}^{LO} (x_1,x_2,z) \dh (z) 
\nonumber \\
&&+\als 
f_{a}^{\H1}(x_1) C_{b}^{\H2}(x_2) d\sigma_{abh}^{LO} (x_1,x_2,z) \dh (z) 
\nonumber \\
&&+\als 
f_{a}^{\H1}(x_1) f_b^{\H2}(x_2) d\sigma_{abh}^{LO}(x_1,x_2,z)  T_{h}^\H(z) + O(\als^2)
\Bigg] dx_1 dx_2 dz\,,
\label{mastereqn}
\eeqn
where all the matrix elements in the partonic cross sections 
($d\sigma_{abh}^{LO}$, $d\sigma_{abh}^{NLO}$)
are crossed versions of the ones with all resolved partons in the final state.

This completes the treatment of the massless fragmentation functions.
We next extend the PSS formalism to include heavy quarks.

\section{Massive Particles}

A very important class of tags consists of heavy (D or B) mesons,
indicating that a heavy quark was produced in the hard scattering. A
description of such reactions with massless quarks is often not
sufficiently accurate.  In this section, we show how to
extend the previous results to cases involving heavy quarks with an
explicit mass.  Throughout this section the mass is understood to be
renormalized in the on-mass-shell scheme.

\subsection{Soft Contribution}

We consider the reaction:
\beq
\Big(l \bar{l'} \go\Big) V \go Q_1 \overline{Q}_2 + (\ng+1)g\,,
\label{vecdecay}
\eeq
where $V$ stands again for any vector boson (e.g. $\gamma, Z, W$, etc).  
Now $Q_1$ and $Q_2$ are heavy quarks with masses
$m_1$ and $m_2$, respectively.  Our results can be straightforwardly
extended to the case with four massive quarks $V \go Q_1\bar{Q}_2
Q_3\bar{Q}_4 +(\ng-1)g$, or two heavy and two light ones, etc.

When gluon $s$ become soft, the colorless subamplitude $S_\mu$
again factorizes into an eikonal factor multiplying the ordered subamplitude 
with one less gluon.
The square of the eikonal factor is, summed over the helicities of the soft gluon,
\beq
f_{ab}(s) =  \sum_{\lambda}\left| e_\lambda(a;s;b) \right|^2 =\frac{4s_{ab}}{s_{as} s_{bs}}
- \frac{4m_a^2}{s_{as}^2} -\frac{4m_b^2}{s_{bs}^2}\,,
\eeq  
where we keep the definition of $s_{as}$ as in section 2:
\beq
s_{as}\equiv 2 p_a\cdot p_s\,,
\eeq
and $m_a$ and $m_b$ are the masses of the neighboring particles, which can be
$m_1$, $m_2$, or zero.
To analyse the soft behavior of phase space, we only need to understand the three particle
case, without loss of generality. The three particle phase space measure
reads in $d=4-2\eps$ dimensions 
\beq
dP^d(\Q;a,b,s)= (2\pi)^{3-2d}
\frac{d^{d-1}p_a}{2E_a}
\frac{d^{d-1}p_b}{2E_b}
\frac{d^{d-1}p_s}{2E_s}
\delta^d (\Q-p_a-p_b-p_s) 
\eeq
The soft region is defined as in the massless case: $s_{as}, s_{bs}
<\smin$, and $\smin$ is assumed to be much smaller than $\Q^2$.  
This phase space
measure again factorizes, up to terms of order $\smin/\Q^2$: 
\beq
dP^d(\Q;a,b,s) \go dP^d(\Q;a,b)\; dP^{\eps}_{soft}(a,b,s)\,.  
\eeq 
The soft gluon measure
\beq dP^{\eps}_{soft}(a,b,s) = (2\pi)^{1-d}\frac{d^{d-1}p_s}{2E_s}
\theta(\smin-s_{as}) \theta(\smin-s_{bs}) 
\eeq 
may be written as 
\beqn
dP^{\eps}_{soft}(a,b,s) &=& \frac{(4\pi)^\eps}{16\pi^2}
\frac{\lambda^{(\eps -\half)}}{\Gamma (1-\eps)} \left[ s_{as} s_{bs}
s_{ab} - m_b^2 s_{ab}^2- m_a^2 s_{bs}^2 \right]^{-\eps} \times \nonumber \\ 
&&ds_{as} ds_{bs} \theta (\smin-s_{as}) \theta(\smin-s_{bs})\,, 
\eeqn 
with
\beq 
\lambda = s_{ab}^2 -4\,m_a^2 \, m_b^2\,.  
\eeq 

Note that if one takes $m_a$ and $m_b$ to zero, one recovers the massless result
in Eq.~(\ref{softps}).  The factor between square brackets is in fact
proportional to $(p_s^{\perp})^2$ , where $p_s^{\perp}$ is the
component of $p_s$ in the direction perpendicular to $p_a$ in the
center of mass frame of $a$,$b$, and $s$, and must hence be positive.  
This positivity condition leads to the kinematical
constraint~\footnote{This result can also be derived from the
condition that $|\cos(\theta)|\leq 1$, where $\theta$ is the angle
between $a$ and $s$ in the center of mass frame of $a$,$b$, and $s$.}  
\beq
\tau_- s_{bs} <s_{as} < \tau_+ s_{bs}\,, \eeq or equivalently \beq
\sigma_- s_{as} <s_{bs} < \sigma_+ s_{bs}, \eeq where \beq \tau_{\pm}
= \frac{s_{ab}}{2m_b^2}\pm\sqrt{(\frac{s_{ab}}{2m_b^2})^2
-\frac{m_a^2}{m_b^2}} \;,\;\;\; \sigma_{\pm}
=\frac{s_{ab}}{2m_a^2}\pm\sqrt{(\frac{s_{ab}}{2m_a^2})^2
-\frac{m_b^2}{m_a^2}}.  \eeq Note that
$(\tau_{\pm})^{-1}=\sigma_{\mp}$.  For the rest of this discussion we
assume $m_a \geq m_b$.

One can show that $\tau_- \leq(\geq)\; 1 $ if $s_{ab} \geq (\leq)\; m_a^2 +m_b^2$.  
Furthermore $\tau_+ > 1$.
The complete integration region for $s_{as}$ and $s_{bs}$
is conveniently divided into two regions. The first region is 
\beq
s_{bs}\in [0,\sigma_- \smin]\;, \;\;\; 
s_{as}\in [\tau_- s_{bs},\tau_+ s_{bs}]\, ,
\label{rangeone}
\eeq
and the second region is
\beqn
&1)& s_{ab} \geq m_a^2 +m_b^2: \;\;\;
s_{bs}\in [\sigma_- \smin,\smin]\;, 
  \;\; s_{as}\in [\tau_- s_{bs},\smin]\,,\nonumber \\
&2)& s_{ab} \leq m_a^2 +m_b^2: \;\;\;
s_{bs}\in [\sigma_- \smin,\sigma_+ \smin]\;, 
  \;\; s_{as}\in [\tau_- s_{bs},\smin]\, .
\label{rangetwo}
\eeqn
The threshold condition is $s_{ab} \geq 2\,m_a m_b$.
Note that $m_b=0$ implies $\sigma_-=0$, in which case the first
region of integration does not contribute, and in the second
region the lower limit of $s_{bs}$ reduces to zero.

Combining the soft part of the phase space and matrix element
in this general massive case, we find 
\beqn
 S(m_a,m_b) &=& \int \left( \frac{g^2 N}{2} \right) f_{ab}(s) dP^{\eps}_{soft} (a,b,s) . 
\label{softint}
\eeqn
We distinguish between two cases in the rest of this section,
because an extra divergence occurs when $m_b=0$.  We find the
following results after performing the integration in Eq.~(\ref{softint}): 
\subsubsection*{$m_b=0$ case}
Denoting $m_a\equiv m$ we find 
\beq
S(m,0)= \frac{\als N}{2\pi} \frac{1}{\Gamma (1-\eps)} 
\left( \frac{4\pi \mu^2}{\smin} \right)^\eps 
\left( \frac{s_{ab}}{\smin} \right)^\eps  J(m,0)
\eeq
where, corresponding to the two ranges in (\ref{rangetwo})
\beq
1)\;\; s_{ab} \geq m^2 \;:\;\;\;
 J(m,0)= \frac{1}{\eps^2} 
- \frac{1}{2\eps^2}\left( \frac{s_{ab}}{m^2}\right)^\eps
+\frac{1}{2\eps} \left( \frac{s_{ab}}{m^2} \right)^\eps
-\frac{1}{2} \zeta (2) + \frac{m^2}{s_{ab}}
\nonumber\eeq
\beq
2)\;\; s_{ab} \leq m^2 \;:\;\;\;
J(m,0)= \left( \frac{s_{ab}}{m^2} \right)^{-\eps}
\left( \frac{1}{2 \eps^2} 
+\frac{1}{2\eps} 
-\frac{1}{2} \zeta (2) + 1 \right)
\label{eq:jm} 
\eeq
Note that the strength of the $1/\epsilon^2$ poles is the same in both cases.  

\subsubsection*{$m_b\neq 0$ case}
Here we find
\beqn
S(m_a,m_b)= \frac{\als N}{2\pi} \frac{1}{\Gamma (1-\eps)} 
\left( \frac{4\pi \mu^2}{\smin} \right)^\eps 
\frac{(m_b^2)^{1-2\eps}}{\lambda^{\half -\eps}}
\left( \frac{m_b^2}{\smin} \right)^\eps  J(m_a,m_b) .
\eeqn
The contribution to $J$ coming from the first region of integration 
(\ref{rangeone}) has the following form:
\beqn
J(m_a,m_b)&=& -\frac{\tau_+^{2\eps}}{2\eps} \left( 2(\tau_--\tau_+) 
+ (\tau_-+\tau_+) \ln(\frac{\tau_+}{\tau_-}) \right)
-2 (\tau_+-\tau_-) \ln(\tau_+-\tau_-) \nonumber \\
&&+(\tau_+-\tau_-)+\frac{\tau_-+\tau_+}{2} 
\ln(\frac{\tau_+}{\tau_-})[1+2 \ln(\tau_+-\tau_-)]
\nonumber \\
&&
+\Big(\frac{\tau_-+\tau_+}{2}\Big) \Big({\rm Li}_2(1-\frac{\tau_+}{\tau_-}) 
- {\rm Li}_2(1-\frac{\tau_-}{\tau_+})\Big).
\eeqn
The contribution from the second region of integration (\ref{rangetwo}) is
\beqn
&1)& s_{ab} \geq m_a^2+m_b^2:\;\;\;
 J= 1 -(\tau_-+\tau_+) + \tau_-\tau_+ \nonumber \\
&& + \ln(\tau_+) (\tau_--\tau_+-(\tau_-+\tau_+) 
\ln(\tau_-)) + \half (\tau_-+\tau_+) \ln^2(\tau_+) \\
&2) & s_{ab} \leq m_a^2+m_b^2:\;\;\;
J= (\tau_--\tau_+) \ln(\frac{\tau_+}{\tau_-}) 
+\frac{(\tau_-+\tau_+)}{2} \ln^2(\frac{\tau_+}{\tau_-}) \,.
\eeqn

We cannot take the limit $m_b\rightarrow 0$ in this expression, which
reflects the extra singularity that leads to a double pole in
Eq.~(\ref{eq:jm}).  Note finally that if $m_a=m_b=m$ there is only the
case $s_{ab} \geq 2m^2$ (and $\sigma_{\pm} =\tau_{\pm}$), because the
other case violates the threshold condition.

It is obvious how to modify the soft factor $S_F$ in Eq.~(\ref{softfac}) 
for reaction (\ref{vecdecay}):
\beqn
&n_g > 0&:\;\;\; S_F(Q_1;1,\ldots,\ng;\overline{Q_2}) = 
\Big(\frac{\alpha_s N}{2 \pi}\Big)\frac{1}{\Gamma(1-\epsilon)}
\Big(\frac{4\pi\mu^2}{\smin}\Big)^\eps \\
&\times&\Big[J(m_1,0)\Big(\frac{s_{K1}}{\smin}\Big)^{\epsilon}
+\frac{1}{\epsilon^2}\Big(\frac{s_{12}}{\smin}\Big)^{\epsilon}
+\ldots+
\frac{1}{\epsilon^2}\Big(\frac{s_{(\ng-1)1}}{\smin}\Big)^{\epsilon}
+J(0,m_2)\Big(\frac{s_{\ng\overline{K}}}{\smin}\Big)^{\epsilon}\Big] \nonumber \\
&n_g = 0&:\;\;\; S_F(Q_1;\overline{Q_2}) = 
\Big(\frac{\alpha_s}{2 \pi}\Big)\Big(\frac{N^2-1}{N}\Big)\frac{1}{\Gamma(1-\epsilon)}
\Big(\frac{4\pi\mu^2}{\smin}\Big)^\eps \nonumber\\
&\times&\Big[\frac{(m_2^2)^{1-2\eps}}{\lambda^{1/2-\eps}}
\Big(\frac{m_2^2}{\smin}\Big)^\eps J(m_1,m_2)\Big]\, .
\label{skfmass}
\eeqn
Note the replacement of $N$ by $(N^2-1)/N$ in the case $\ng=0$.

There are no $1/\epsilon$ singularities 
associated with the collinear region when
one of the two collinear particles is a heavy quark.  The mass of the
quark screens the singularity.  Therefore the above extension of the
soft $K$-factor to the massive case is all that is needed to 
be able to apply the PSS method to processes involving heavy quarks.
As far as UV renormalization is concerned in this situation, any
scheme is of course allowed, but the Collins-Wilczek-Zee~\cite{CWZ}
scheme, in which the decoupling of the heavy quarks for small external
momenta is manifest, is preferred.  In this case only the light quarks
contribute to the running of the QCD coupling.

The PSS method, extended as above to include soft radiation
from heavy quarks, has in the recent past already been used in 
constructing NLO programs for reactions where heavy quarks are produced in 
leptonic collisions \cite{PSSmassother}, and also in
hadronic collisions using crossing functions \cite{GKL1}. We 
shall next extend it further to include collinear radiation.

\subsection{Collinear Behavior and Massless Limit}

In a scattering process involving a heavy quark in which the typical hard scale $\Q$ 
(e.g. the $p_T$ of the heavy quark) is
of the order of the heavy quark mass $m$, the results obtained thus
far are adequate to describe the reaction.  Singularities due to collinear 
configurations involving the heavy quark are screened by its mass, 
logarithms $\ln(\Q/m)$ are small, and the cross
section is well-behaved.  
However if $\Q/m$ is large, the collinear region contributes large
logarithms $\ln(\Q/m)$ that dominate the cross section at higher
orders in perturbation theory. Moreover, they may spoil the convergence of
the QCD perturbation series and one may wish to resum them.
The $\ln(m)$ terms are the equivalent of the $1/\eps$ poles that appear 
in the collinear region with massless partons in dimensional regularization
and can be treated conceptually on equal footing. In what follows
we describe how heavy quarks can be incorporated into the PSS formalism
if $\Q \gg m$.

The outline of this subsection is as follows.
We first present results for the behavior of the cross section in the collinear region 
when the heavy quark is in the final state in the 
limit that $m/\Q$ and $m^2/\smin$ are small.  Then, we derive the
heavy quark contribution to the tagging functions. 
We present results for two factorization schemes, 
a minimal substraction (``\ms'') and massive \msb (``\mmsb'') scheme, 
the latter defined such that the heavy quark 
fragmentation functions are the same as in the \msb massless case.
We show that the counterterm  functions of the massive \msb scheme are in fact 
equal to the perturbative fragmentation functions of
Ref.~\cite{MN}.
The fragmentation functions absorb in this way, through factorization, the $\ln(m)$ terms
and resum the $\ln(\Q/m)$ terms through their evolution.  
We then present the equivalent results for the initial state. 

A comprehensive treatment of heavy
quark cross sections should be performed in the context
of a  variable flavor number
scheme~\cite{ACOT94}, which aims to describe the
full region from $\Q\sim m$ to $\Q \gg m$. This subject has
recently been an active topic of 
investigation \cite{Buza,LaiTung,MRRS,ThorneRoberts,Collins,Scalise}.
The results presented in this paper can be used to implement this scheme and
we comment on this possibility at the end of this section.

\subsubsection*{Final State}

We first derive the behavior of the cross section in the collinear
limit when the typical hard scale $\Q$ is much larger than $m$.  
As in the massless case, we take all partons and massive 
quarks as outgoing.  Because the mass serves as a regulator for the
collinear divergences, we may work in four rather than in
$d$ dimensions.
We follow the structure of section 2.2.  
Let us denote the particles that become collinear by $A$ and $B$, 
and their parent particle by $C$,
\beqn \label{eq:hqcoldef}
p_C&=&p_A + p_B\; ;\qquad p_A = z p_C\; ; \qquad p_B = (1-z)p_C \nonumber \\ 
\delta_{AB}&=&p_C^2-m_C^2=(p_A+p_B)^2-m_C^2 < \smin.  
\eeqn 
$\delta_{AB}$ is the inverse propagator of $C$.
The capital letters for the momenta remind us that two of the 
$A$, $B$ and $C$ particles are heavy quarks.  
We have defined $z$ in Eq.~(\ref{eq:hqcoldef}) as a 4-momentum
fraction. 
Other definitions (such as an energy, or 3-momentum fraction) 
are equivalent in the collinear region when $m \ll \Q$.
As usual we assume that
$\smin$ is small compared to $\Q^2$.  We furthermore neglect terms 
that give contributions of the order of $m^2 /\smin$ to the 
cross section. 

As in the massless case, in the collinear region
the matrix elements factorize:
\beq
| V^{\mu}\hat{S}_{\mu}(1,\ldots,A,B,\ldots,n+1)|^2 \go
\hat{c}_F^{C\go AB} | V^{\mu}\hat{S}_{\mu}(1,\ldots,C,\ldots,n)|^2\ ,
\eeq
with
\beq \label{eq:cffsdef}
\hat{c}_F^{C\go AB} = \Big(\frac{g^2N}{2}\Big) f^{C\go AB}
\eeq
and
\beq
f^{C\go AB}=\frac{P^{m}_{C\go AB}(z,m,\delta_{AB})}{\delta_{AB}} \, .
\eeq
The massive splitting functions $P^{m}_{C\go AB}$ contain
a term involving the mass $m$ of the heavy quark ($Q$) and $\delta_{AB}$:
\begin{equation}
P^{m}_{Q\rightarrow Qg}(z,m,\delta_{Qg})
=2(1-\frac{1}{N^2}) \Big[ 
\Big(\frac{1+z^2}{1-z}\Big)-\frac{2m^2}{\delta_{Qg}}\Big] \,,
\label{Pmone}
\end{equation}
\begin{equation}
P^{m}_{Q\rightarrow gQ}(z,m,\delta_{Qg})
=2(1-\frac{1}{N^2}) \Big[ 
\Big(\frac{1+(1-z)^2}{z}\Big)-\frac{2m^2}{\delta_{Qg}}\Big] \,,
\label{Pmtwo}
\end{equation}
where $\delta_{Qg}= 2p_Q\cdot p_g(\equiv s_{Qg})$, and
\begin{equation}
P^{m}_{g\rightarrow Q\bar{Q}}(z,m,\delta_{Q\bar{Q}})
=\frac{2}{N}\Big[ z^2+(1-z)^2+
\frac{2m^2}{\delta_{Q\bar{Q}}}\Big] 
\label{Pmthree}
\end{equation}
where $\delta_{Q\bar{Q}} = (p_Q+p_{\bar{Q}})^2$.  
The terms containing the mass can be calculated by the methods used in~\cite{AP}.

The phase space also factorizes in the collinear region:
\beq
dP^4(\Q;1,\ldots,n-1,A,B) \rightarrow dP^4(\Q;1,\ldots,n-1,C)\ 
dP_{coll}(A,B;z)\ , 
\eeq
where
\beq
dP_{coll}(A,B;z)=\frac{1}{16\pi^2} dz d\delta_{AB}.
\eeq
The integrated collinear factor is then given by:
\begin{eqnarray}
\int  \Big(\frac{g^2N}{2}\Big) f^{C\go AB}  dP_{coll}(A,B;z)
&=& \Big(\frac{\alpha_s N}{2 \pi}\Big) I_{C\go AB}(z_1,z_2)\, ,
\end{eqnarray}
where
\begin{eqnarray}
I_{C\go AB}(z_1,z_2)=\frac{1}{4}
\int_{z_1}^{z_2} dz \int_{\delta_{\rm min}}^{\smin}
\frac{d\delta_{AB}}{\delta_{AB}}  
P^{m}_{C\go AB}(z,m,\delta_{AB})
\end{eqnarray}
$z_1$ and $z_2$ are defined as in the massless case, and are needed to
avoid the soft region when $A$ or $B$ is a gluon.
The kinematic limits $\delta_{\rm min}$ are different for each of the splittings
and are given in Table~\ref{tab:delta}.
\begin{table}[ht]
\caption{The kinematic limits $\delta_{\rm min}$ for tagging functions.}
\label{tab:delta}
\begin{center}
\begin{tabular}{|c|c|} \hline
$C\go AB $&  $\delta_{\rm min}$ \\
\hline
$Q\rightarrow Qg $&$\frac{m^2 (1-z)}{z}$ \\
\hline
$Q\rightarrow gQ $& $\frac{m^2 z}{(1-z)}$ \\
\hline
$g\rightarrow Q\bar{Q} $&$  \frac{m^2}{ z(1-z)}$ \\ 
\hline
\end{tabular}
\end{center}
\end{table}
Note that although $m^2$ is small compared to $\smin$, we must retain it
in $\delta_{\rm min}$, because it will give rise to $\ln{(m^2/\smin)}$ 
contributions to the cross section that cannot be neglected.  
Other mass terms in the limits of integration
that give rise to order $(m^2/\smin)$ contributions have been neglected.
We define at this point the $K_{C \go A}$ functions via:
\begin{equation}
I_{C\go AB}(z_1,z_2)=\int_{z_1}^1 dz K_{C \go A} (z,z_2).
\end{equation}
When $z_2$ cannot be put to one (when $B$ is a gluon) 
its effect is absorbed in the integrand.
With the plus prescription defined in Eq.~(\ref{plusdef}), we obtain the 
following results:
\beqn
K_{Q\go Q}(z,z_2) &=& \frac{1}{2}(1-\frac{1}{N^2})
\Bigg[\frac{1}{(1-z)_+}\Big((1+z^2)\ln\Bigg(\frac{\smin}{m^2}z\Bigg)-2z\Big)
 \\ &-&(1+z^2)\Big(\frac{\ln(1-z)}{1-z}\Big)_+ 
-\delta(1-z)\Bigg(2\ln z_2\Big(\ln\Bigg(\frac{\smin}{m^2}\Bigg)-1\Big)
-\ln^2z_2\Bigg)\Bigg]\,,
\nonumber
\eeqn
\beq
K_{Q\go g}(z,0) = \frac{1}{2}(1-\frac{1}{N^2})
\Bigg[\frac{1+(1-z)^2}{z}\ln\Big(\frac{\smin}{m^2}\frac{1-z}{z}\Big)
-\frac{2(1-z)}{z}\Bigg]\,,
\eeq
\beq
K_{g\go Q}(z,0) = 
\frac{1}{2N}\Bigg[(z^2+(1-z)^2)\ln\Bigg(\frac{\smin}{m^2}z(1-z)\Bigg)
+2z(1-z)\Bigg]\,.
\eeq
Consequently,
\beqn
I_{g\rightarrow Q\bar{Q}}(0,0) &=&
\frac{1}{2N}
\int_0^1 dz
\Bigg[(z^2+(1-z)^2)\ln\Big(\frac{\smin}{m^2}z(1-z)\Big)
+2z(1-z)\Bigg] \nonumber \\
&=& \frac{1}{2N} \left(\frac{2}{3}\ln\Big(\frac{\smin}{m^2}\Big)-\frac{10}{9}\right)\,,
\label{eq:firstIm}
\eeqn 
\beqn
I_{Q\rightarrow Qg}(0,z_2) &=&
\frac{1}{2}(1-\frac{1}{N^2})
\int_0^{1-z_2} dz
\Bigg[(\frac{1+z^2}{1-z})\ln\Big(\frac{\smin}{m^2}\frac{z}{1-z}\Big)-\frac{2z}{1-z}\Bigg] \\
&=& \frac{1}{2}(1-\frac{1}{N^2})\Bigg[(-2\ln z_2-\frac{3}{2})\ln\Big(\frac{\smin}{m^2}\Big)
+\ln^2z_2+2\ln z_2 -\frac{\pi^2}{3}+\frac{3}{2}\Bigg] \,,\nonumber
\label{eq:secondIm}
\eeqn
and
\beqn
I_{Q\rightarrow gQ}(z_1,0) &=& I_{Q\rightarrow Qg}(0,z_1).
\label{eq:thirdIm}
\eeqn
The massive $I_{C \go AB}$ functions can be used, as in the massless case,
to calculate the collinear factor $C_F(\ldots)$ (see Eq.~(\ref{eq:cf})) 
for each of the colorless subamplitudes.  
For sufficiently inclusive, infrared-safe observables, the $\ln(m^2)$ terms 
of the $I_{C \go AB}$ functions will cancel with corresponding terms of the virtual 
contribution, the same way the $1/\eps$ terms cancel in the massless case.  
In these cases it is better to integrate the collinear region analytically
as done here and perform the cancellations analytically.  
The numerical integration and cancellation, although always possible, is
in practice difficult at large $\Q$.

We now turn to the problem of including the heavy quark into
our formalism for fragmentation functions. 
For this we must add the heavy quark contribution to
the tagging functions, where we assume that the transition
of the heavy quark to a heavy hadron,
at least for values of $\Q$ much larger than $m$,
may be described by convoluting the partonic matrix
element with a (non-perturbative) fragmentation function.
In the normalization conventions of this section the heavy quark contribution 
to the tagging functions can be constructed via 
\beq
T_{h}^\H (z)=T_{H,conv}^\H (z)-T_{H,int}^\H(z) \equiv\sum_p \int_{z}^{1}
\frac{dy}{y} Y_{H \go P}(y,m) D^{\H}_P(z/y)\, , \eeq where 
\beqn
T_{H,conv}^{\H}(z) &=& \facm \sum_p \int_z^1 {dy\over y} K_{H\go
P}(y,y_2) D_P^{\H}(\frac{z}{y})\,, \eeqn and 
\beq T_{H,int}^\H(z) =
\frac{N}{2\pi}\sum_{P}I_{H\go PU}(y_1,y_2) D_H^{\H}(z)\,.  \eeq
\H can now be a light or heavy hadron and $H$, $P$, and $U$ can be heavy quarks.

There are four possible contributions due to the heavy quark:
\beqn
Y_{Q\go Q}(y,m) &=& \facm\Bigg(
K_{Q\go Q}(y,y_2)-\delta(1-y)I_{Q\rightarrow Qg}(0,y_2)\Bigg) \nonumber \\
 &=& \facm\frac{1}{2}(1-\frac{1}{N^2})\Bigg[\Big(\frac{1+y^2}{1-y}\Big)_+\ln\Big(\frac{\smin}{m^2}\Big)
+\frac{(1+y^2)\ln y}{(1-y)_+} \nonumber \\
&-& \frac{2y}{(1-y)_+}-(1+y^2)\Big(\frac{\ln(1-y)}{1-y}\Big)_+
+(\frac{\pi^2}{3}-\frac{3}{2})\delta(1-y)\Bigg]\,,\\
Y_{g\go Q}(y,m) &=& \facm K_{g\go Q}(y)\,, \\
Y_{Q\go g}(y,m) &=&\facm K_{Q\go g}(y) \,,\\
\eeqn
and
\beq
Y^m_{g\go g}(y,m) = -\facm I_{g\rightarrow Q\bar{Q}}(0,0) \delta(1-y)\,.
\label{yqgg}
\eeq
$Y^m_{g\go g}$ is the heavy quark contribution that must be added to the massless 
$Y_{g\go g}$.
As in the massless case, all dependence on the boundary $y_2$ has
cancelled. Eq.~(\ref{eq:sh2}) can now also be applied to the massive
fermion case.  The heavy quark contribution to the tagging functions
is independent of the hard process. 

When $ln(\Q/m)$ becomes too large we need to these logarithms
to all orders.  This can be done by the evolution of the fragmentation functions,
through factorization.  The factorization procedure, as in Eq.~(\ref{taggconstr}), leads to
scheme dependent $Y_{H\go P}$ functions: 
\beq 
Y_{H\go P}(y,m) \go
Y^{\rm scheme}_{H\go P}(y,\mu_D) = Y_{H\go P}(y,m) + R^{\rm
scheme}_{H\go P}(y,m,\mu_D)\,.
\label{mfone}
\eeq
A minimal substraction, ``${\rm MS}$", choice of the counterterm functions
$R_{H\go P}$ is:
\beq
R^{\rm MS}_{Q\go Q}(y,m,\mu_D) = \facm 
\frac{1}{2}(1-\frac{1}{N^2}) \Bigg[\Big(\frac{1+y^2}{1-y}\Big)\Bigg]_+ 
\ln(\frac{m^2}{\mu_D^2})\,,
\label{rone}
\eeq
\beq
R^{\rm MS}_{Q\go g}(y,m,\mu_D) = \facm \frac{1}{4} P^{\eps=0}_{q\go gq}(y) \ln(\frac{m^2}{\mu_D^2})\,,
\label{rtwo}
\eeq
\beq
R^{\rm MS}_{g\go Q} (y,m,\mu_D)= \facm \frac{1}{4} P^{\eps=0}_{g\go q\bar{q}}(y) 
\ln(\frac{m^2}{\mu_D^2})\,,
\label{rthree}
\eeq 
\beq
R^{m,{\rm MS}}_{g\go g}(y,m,\mu_D) = -\facm 
\frac{2}{6N} \ln(\frac{m^2}{\mu_D^2})\delta(1-y) \,.
\label{rfour}
\eeq
where the $P^{\eps=0}_{c\go ab}$ can be inferred from 
the massless splitting functions in Eq.~(\ref{eq:Peps}).  
$R^m_{g\go g}$ is the massive contribution that must be added to the 
massless $R_{g\go g}$.  
Notice that the heavy flavor is active in the evolution of the
fragmentation function.
The $Y^{{\rm MS}}_{H\go P}$ are now finite for 
$m\go 0$ at fixed $\mu_D$.

We can decompose the heavy quark contribution to the tagging functions 
as in Eq.~(\ref{eq:UVdef}) for the massless case, with:  
\beq
U_{Q \go Q}^{\H}(z,\mu_D)=\frac{1}{2}(1-\frac{1}{N^2})\int_z^1 \frac{dy}{y} 
 \Bigg[\Big(\frac{1+y^2}{1-y}\Big)\Bigg]_+ D_Q^{\H}(z/y,\mu_D)\,,
\label{umfirst}
\eeq
\beq
U_{g \go Q}^{\H}(z,\mu_D)=\frac{1}{4} \int_z^1 \frac{dy}{y} 
P^{\eps=0}_{g \go q\bar{q}}(y) 
D_Q^{\H}(z/y,\mu_D)\,,
\eeq
\beq
U_{Q \go g}^{\H}(z,\mu_D)=\frac{1}{4} \int_z^1 \frac{dy}{y} 
P^{\eps=0}_{q \go g q}(y) 
 D_g^{\H}(z/y,\mu_D)\, ,
\eeq
\beq
U^{m,\H}_{g \go g}(z,\mu_D)=-\frac{2}{6N} D_g^{\H}(z,\mu_D)  
\eeq
\beqn
V_{Q \go Q}^{\H,{\rm MS}}(z,\mu_D)=\frac{1}{2} (1-\frac{1}{N^2})
\int_z^1 \frac{dy}{y} 
\Bigg[
\ln y \frac{1+y^2}{(1-y)_+}
+(\frac{\pi^2}{3} -\frac{3}{2}) \delta (1-y)
\nonumber \\
-\frac{2y}{(1-y)_+} -(1+y^2) \Bigg(\frac{\ln(1-y)}{1-y}\Bigg)_+
\Bigg] D_Q^{\H}(z/y,\mu_D)\,,
\eeqn
\beq
V_{g \go Q}^{\H,{\rm MS}}(z,\mu_D)=\frac{1}{2N} \int_z^1 \frac{dy}{y} 
\Bigg[
\Big(y^2+(1-y)^2\Big)\,\ln(y(1-y)) + 2y(1-y)
\Bigg]\,D_Q^{\H}(z/y,\mu_D)\,,
\eeq
and
\beq
V_{Q \go g}^{\H,{\rm MS}}(z,\mu_D)= \frac{1}{2} (1-\frac{1}{N^2})\int_z^1 \frac{dy}{y} 
\Bigg[
\frac{1+(1-y)^2}{y}\ln\Big(\frac{1-y}{y}\Big)-\frac{2(1-y)}{y}
\Bigg]\,D_g^{\H}(z/y,\mu_D)\,.
\label{vmlast}
\eeq
\beq
V^{m,\H,{\rm MS}}_{g \go g}(z,\mu_D)=\frac{5}{9N} D_g^{\H}(z,\mu_D)  
\eeq

The $U^{\H}$ are scheme independent, and have the same functional dependence
on their respective fragmentation functions as in the massless case.
The contribution of the heavy quark to $U_{g \go g}$ and $V_{g \go g}$ 
increases the number of flavors $n_f$ in the massless expressions by one.  
After factorization the heavy quark contribution to the tagging functions
is independent of $m$ and is therefore finite as $m$ goes to zero.

However, the tagging functions, although now $m$-independent, 
are not the same as those corresponding to the massless case of 
section 3. This implies that in the \ms scheme the $m\go 0$
limit of the full partonic cross section, which can safely be
taken, is not the same as if the calculation had been done with
massless partons, and in the massless \msb scheme, from the outset. When $\Q \gg m$, 
the most appropriate factorization scheme is therefore one 
where the subtracted tagging functions are the same as 
in the $\overline{\rm MS}$ massless scheme,
up to small terms of order $m^2/\Q^2$.  In other words, 
the counterterms in this massive \msb (``\mmsb'') scheme must be such that 
$Y^{\mmsb}_{H\go P}(z,\mu_D)$
is equal to the corresponding massless function
$Y^{\msb}_{h\go p}(z,\mu_D)$. From this
requirement we can straightforwardly infer from our earlier results
what these \mmsb counterterm functions must be, and 
we find
\beqn
R^{\rm \overline{MMS}}_{Q\go Q}(y,m,\mu_D) &=& \facm 
\frac{1}{2}(1-\frac{1}{N^2}) 
\Bigg[
\Big(\frac{1+y^2}{1-y}\Big)_+ \ln(\frac{m^2}{\mu_D^2})
+\frac{1+y^2}{(1-y)_+ } \nonumber \\
&&+ 2 \big( \frac{\ln(1-y)}{1-y}\Big)_+ (1+y^2)
-2 \delta(1-y)
\Bigg]\,,
\label{eq:roneb}
\eeqn
\beq
R^{\mmsb}_{Q\go g}(y,m,\mu_D) = \facm \frac{1}{4} 
P_{q\go gq}(y) \left(\ln\left(\frac{m^2 y^2}{\mu_D^2}\right) +1\right)
\,,
\label{eq:rtwob}
\eeq
\beq
R^{\mmsb}_{g\go Q} (y,m,\mu_D)= R^{\rm MS}_{g\go Q} (y,m,\mu_D)
\label{eq:rthreeb}
\eeq 
\beq
R^{m,{\mmsb}}_{g\go g}(y,m,\mu_D) = R^{m,\rm MS}_{g\go g}(y,m,\mu_D) 
\label{eq:rfourb}
\eeq
The choice of scheme 
has in turn implications for the fragmentation functions, 
which absorb these counterterm functions. They must be determined
in other reactions in the same scheme. 
Notice from the above explicit expressions for the counterterms
that the heavy quark contribution to the kernels for the
evolution equations of the fragmentation functions are still those 
of the massless splitting functions.

We remark that when the contribution from 
the collinear region is calculated analytically in 
the \mmsb scheme as described in this section, the massive
counterterm functions are actually not needed.  All that is needed
to actually perform a calculation, besides the \mmsb fragmentation functions,
are the tagging functions, and in
the \mmsb scheme they are known by definition from the massless \msb case.
However, when the contribution from the collinear region 
and the convolution with the \mmsb fragmentation functions are computed
numerically, then it is necessary to explicitly add the counterterm functions 
in Eqs.~(\ref{eq:roneb}-\ref{eq:rfourb}) 
to make the calculation consistent.
In that case the equivalent tagging functions that should
be used in Eq.~(\ref{eq:sh2}) are given by:
\beq \label{ronly}
\sum_P \int_{z}^{1} \frac{dy}{y} R^{\mmsb}_{H\go P}(y,m,\mu_D) 
D^{\H}_P(z/y,\mu_D)\, .
\eeq    

The original approach towards resumming the large logarithms 
in heavy quark production at high $\Q^2$ is due to
Mele and Nason in Ref.~\cite{MN}, and is called the 
heavy quark perturbative fragmentation functions (PFF) approach.  
Here the large logarithms $\ln(\Q/m)$ are summed at the
partonic level. The resummation again occurs via evolution
of the following initial distributions \cite{MN} at scale $\mu_0$
(of order $m$) to the large scale $\Q$
\beq
D_Q(z,\mu_0,m) = \delta(1-z)+\frac{\alpha_s(\mu_0^2)}{2\pi} d_Q^{(1)} (z,\mu_0,m),
\eeq
and
\beq
D_g(z,\mu_0,m) = \frac{\alpha_s(\mu_0^2)}{2\pi} d_g^{(1)} (z,\mu_0,m).
\eeq
In our formalism the $d_H^{(1)}$ are actually precisely given by the counterterm functions 
(multiplied by $-2\pi$) given in Eqs.~(\ref{eq:roneb}-\ref{eq:rfourb}),
at $\mu_D=\mu_0$.
The expressions in Eqs.~(\ref{eq:roneb}) and (\ref{eq:rthreeb}) 
may be rewritten to 
obtain the same result as in Ref.~\cite{MN}: 
\beq
d_Q^{(1)} (z,\mu_0,m) = \frac{N^2-1}{2N} 
\Bigg[\Big(\frac{1+z^2}{1-z}\Big)
\Big(\ln(\frac{\mu_0^2}{m^2}) -1 -2\ln(1-z)\Big)\Bigg]_+\,,
\label{PFFone}
\eeq
\beq
d_g^{(1)} (z,\mu_0,m) = \Bigg[z^2+(1-z)^2\Bigg]\ln(\frac{\mu_0^2}{m^2})\,.
\label{PFFtwo}
\eeq
The PFF approach has been used to
obtain resummed transverse momentum spectra for a variety of reactions \cite{PFF,NasonOleari}.

\subsubsection*{Initial State}

In this section we present the results needed to include processes
with heavy quarks in the initial state within the PSS formalism, 
when $\Q \gg m$.  As
in the massless case we assume that the calculation of the squared matrix
element was done with all partons, including the heavy quarks, in the final 
state. The heavy quark contribution from the collinear regions
is computed analytically, as just described, under the assumption 
that $\Q$ is large. We now calculate the heavy quark contribution to 
the crossing functions.  We follow the methods and notation of Ref.~\cite{GGK93}, 
as outlined in section 2.3.  Recall that we must add the contributions from the 
initial state collinear region and subtract the contributions from the final
state collinear region that are part of the analytical calculation but
that do not contribute when the unresolved collinear pairs in the 
colorless amplitudes are
crossed.  As in the final state case, 
we neglect terms of order $m^2 / \smin$. 

The process under consideration may be generically represented as
\beq
P + a\go U + M,
\eeq
where $P$ and $a$ are the initial particles of the hard subprocess and 
$M$ can decay into any number of particles.  Particle $U$ is radiated 
collinearly off particle $P$ (and is not observed):
\beq
P\go H + U\, ,
\eeq
whereas $H$ undergoes a hard scattering with $a$ to produce the 
final state labeled $M$.
Here, up to two of $P$, $H$, and $U$ may be heavy quarks, hence the upper case
notation.
The momentum fraction $z$ is again defined by: $p_H = z\,p_P$, $p_U =
(1-z)p_P$, see Eq.~(\ref{eq:hqcoldef}) and the comments below it.
The inverse propagator of the off-shell particle $H$ is given 
by: $\delta_{PU}=(p_P-p_U)^2-m_H^2$.  
The collinear region is defined by $|\delta_{PU}|
< \smin$.  In this limit the phase space and the flux factor
factorize in the following way: 
\beq
\frac{1}{2 s_{Pa}} dP(P+a\go U + M) \rightarrow dP_{coll}(P,U) 
\times  \frac{1}{2 s_{Ha}} dP(H+a \go M)\,.
\eeq 
Note that $s_{Ha}=z s_{Pa}$, where the $z$ factor is 
included in $dP_{coll}(P,U)$. In our notation~\footnote{
$\lambda(x,y,z) = x^2+y^2+z^2-2xy-2xz-2yz$.}, 
$s_{Pa}=2 p_a.p_P = (p_a+p_P)^2-m_P^2=\lambda^{1/2}((p_a+p_P)^2,m_P^2,0)$, 
where we have assumed that $m_a=0$.  

The collinear phase space measure is given by:
\beq 
dP_{coll}(P,U) =
\frac{1}{16\pi^2} z\,dz d|\delta_{PU}|\,.
\eeq

The full squared matrix element factorizes in the collinear region as
described by Eq.~(\ref{eq:mefact}) in the massless case.  
The $\hat{c}_F^{P\go HU}$ are now given by:
\beq
\hat{c}_F^{P\go HU} = \Big(\frac{g^2 N}{2}\Big) \frac{1}{z}\frac{P^{m}_{P\go HU}
(z,m,\frac{|\delta_{PU}|}{z})}{|\delta_{PU}|}\,.
\label{eq:cfQ}
\eeq 
This result can be obtained by appropriate crossing of the final state 
$\hat{c}_F$-factor in Eq.~(\ref{eq:cffsdef}).  The massive splitting functions are given in 
Eqs.~(\ref{Pmone}-\ref{Pmthree}).
The result for $\hat{c}_F^{g\go Q\bar{Q}}$ in Eq.~(\ref{eq:cfQ}) is also given in 
Ref.~\cite{MRRS}.

The initial state component of the crossing functions is given by 
\beqn 
\alpha_s C_{H,init}^\H (x) &=& \int \sum_{P,U} f_P^\H(y) \hat{c}_F^{P\go HU}
dP_{coll}(P,U) \delta(x-zy) dy dz\,, \nonumber \\ 
C_{H,init}^\H(x) &=&  \facm \sum_{P} \frac{1}{4} \int_x^{1-z_2} \frac{dz}{z} 
\int_{\delta_{\rm min}}^{\smin} \frac{d|\delta_{PU}|}{|\delta_{PU}|}
 f_P^\H(x/z) P^{m}_{P\go HU} (z,m,\frac{|\delta_{PU}|}{z})\,, \nonumber \\
&=&  \facm\sum_{P} \int_x^{1} \frac{dz}{z} f_P^\H(x/z) J_{P\go H}(z,z_2)\,.
\label{cminitdef}
\eeqn
Here $\H$ is an initial state hadron, $f_P^\H(x)$ is the
unrenormalized distribution function of $P$ in $\H$.  $z_2$ is introduced as before
to avoid the soft singularity when $U$ is a gluon.
The kinematic limits $\delta_{\rm min}$ are given in Table~\ref{tab:delta2}.
\begin{table}[ht]
\caption{The kinematic limits $\delta_{\rm min}$ for the crossing functions.}
\label{tab:delta2}
\begin{center}
\begin{tabular}{|c|c|} \hline
$p\go hu $&  $\delta_{\rm min}$ \\
\hline
$Q\rightarrow Qg $&$m^2 (1-z)$ \\
\hline
$Q\rightarrow gQ $& $\frac{m^2 z^2}{(1-z)}$ \\
\hline
$g\rightarrow Q\bar{Q} $&$  \frac{m^2}{(1-z)}$ \\ 
\hline
\end{tabular}
\end{center}
\end{table}

Using Eq.~(\ref{plusdef}), we obtain
\beqn
J_{Q\go Q}(z,z_2) &=& \frac{1}{2}(1-\frac{1}{N^2})
\Bigg[\frac{1}{(1-z)_+}((1+z^2)\ln\Big(\frac{\smin}{m^2}\Big)-2z)
-(1+z^2)\Big(\frac{\ln(1-z)}{1-z}\Big)_+ \nonumber \\
&-&\delta(1-z)\Big(2\ln z_2(\ln\Big(\frac{\smin}{m^2}\Big)-1)-\ln^2z_2\Big)\Bigg]\,,
\nonumber\\
J_{Q\go g}(z,0) &=& \frac{1}{2}(1-\frac{1}{N^2})
\Bigg[\frac{1+(1-z)^2}{z}\ln\Big(\frac{\smin}{m^2}\frac{1-z}{z^2}\Big)
-\frac{2(1-z)}{z}\Bigg] \,,\nonumber \\
J_{g\go Q}(z,0) &=& \frac{1}{2N}\Bigg[(z^2+(1-z)^2)\ln\Big(\frac{\smin}{m^2}(1-z)\Big)
+2z(1-z)\Bigg]\,.
\eeqn
Note that the $J_{P \go H}$ can also be obtained from the final 
state $K_{C \go A}$ by appropriate crossings.

The final state contributions to the crossing functions are given by 
\beq
C^\H_{H,final}(x) = (\frac{N}{2\pi})f_H^{\H}(x)\sum_{P}I_{H\go PU}(z_1,z_2)\,,
\eeq
where the $I_{H\go PU}$ are given in Eq.~(\ref{eq:firstIm}-\ref{eq:thirdIm}).
These contributions have to be subtracted:
\beq
C_H^\H(x) = C_{H,init}^\H(x) - C_{H,final}^\H(x)\,.
\eeq
Defining the functions $X_{P\go H}(z,m)$ by
\beq \label{chqufdef}
C_H^\H(x) = \sum_P \int_x^1 \frac{dz}{z} f_P^\H(x/z) X_{P\go H}(z,m)\, ,
\eeq
the four heavy quark contributions are given by,
\beqn
X_{Q\go Q}(z,m) &=& \facm(J_{Q\go Q}(z,z_2)-\delta(1-z)I_{Q\rightarrow Qg}(0,z_2))\,, \nonumber \\
 &=& \facm\frac{1}{2}(1-\frac{1}{N^2})
\Bigg[\Big(\frac{1+z^2}{1-z}\Big)_+\ln\Big(\frac{\smin}{m^2}\Big)
\nonumber \\
&-& \frac{2z}{(1-z)_+}-(1+z^2)\Big(\frac{\ln(1-z)}{1-z}\Big)_+
+(\frac{\pi^2}{3}-\frac{3}{2})\delta(1-z)\Bigg]\, , \nonumber\\
X_{g\go Q}(z,m) &=& \facm J_{g\go Q}(z)\, ,\nonumber \\
X_{Q\go g}(z,m) &=& \facm J_{Q\go g}(z)\, ,\nonumber \\
X^m_{g\go g}(z,m) &=& -\frac{N}{2\pi} I_{g\rightarrow Q\bar{Q}}(0,0) \delta(1-z)\,.
\label{massX}
\eeqn
where $X^m_{g\go g}$ is the massive quark contribution to the massless
$X_{g\go g}$.

After factorization, using the minimal subtraction scheme defined by the massive 
counterterm functions given in Eq.~(\ref{rone}-\ref{rfour}) (and replacing 
$\mu_D$ by $\mu_F$), the crossing function can be written as in the massless case:
\beq
C_H^{\H,{\rm MS}}(x,m,\mu_F) = \frac{N}{2\pi}\Big[A_H^{\H}(x,\mu_F)\ln(\frac{\smin}{\mu_F^2})
+B_H^{\H,{\rm MS}}(x,\mu_F)\Big]\, ,
\label{cmdefab}
\eeq
with
\beq
A_{H}^{\H}(x)=\sum_P A_{P\go H}^{\H} (x,\mu_F)\;\;,\;\;
B_{H}^{\H,{\rm MS}}(x)=\sum_P B_{P\go H}^{\H} (x,\mu_F) \,,
\eeq
we have
\beq
A_{Q \go Q}^{\H}(x,\mu_F)=\frac{1}{2}(1-\frac{1}{N^2})\int_x^1 \frac{dz}{z} 
f_Q^{\H}(x/z,\mu_F) \Bigg[\Big(\frac{1+z^2}{1-z}\Big)_+\Bigg] \,,
\label{amfirst}
\eeq
\beq
A_{g \go Q}^{\H}(x,\mu_F)=\frac{1}{4} \int_x^1 \frac{dz}{z} 
f_g^{\H}(x/z,\mu_F) P^{\eps=0}_{g\go q \bar{q}} (z)\,,
\eeq
\beq
A_{Q \go g}^{\H}(x,\mu_F)=\frac{1}{4} \int_x^1 \frac{dz}{z} 
 f_Q^{\H}(x/z,\mu_F)  P^{\eps=0}_{q\go g q} (z)\,,
\eeq
\beq
A_{g \go g}^{m,\H}(x,\mu_F)=-\frac{1}{3N} f_g^{\H}(x,\mu_F) \,,
\eeq
and for the scheme dependent functions
\beqn
B_{Q \go Q}^{\H,{\rm MS}}(x,\mu_F)=\frac{1}{2} (1-\frac{1}{N^2})
\int_x^1 \frac{dz}{z} 
f_Q^{\H}(x/z,\mu_F)\Bigg[
(\frac{\pi^2}{3} -\frac{3}{2}) \delta (1-z)
\nonumber \\
-\frac{2z}{(1-z)_+} -(1+z^2) \Bigg(\frac{\ln(1-z)}{1-z}\Bigg)_+
\Bigg] \,,
\eeqn
\beq
B_{g \go Q}^{\H,{\rm MS}}(x,\mu_F)=\frac{1}{2N} \int_x^1 \frac{dz}{z} 
f_g^{\H}(x/z,\mu_F)\Bigg[
(z^2+(1-z)^2)\,\ln(1-z) + 2z(1-z)
\Bigg]\,,
\eeq
\beq
B_{Q \go g}^{\H,{\rm MS}}(x,\mu_F)= \frac{1}{2} (1-\frac{1}{N^2})\int_x^1 \frac{dz}{z} 
f_Q^{\H}(x/z,\mu_F)\Bigg[
\frac{1+(1-z)^2}{z}\ln(\frac{1-z}{z^2})-\frac{2(1-z)}{z}
\Bigg]\,,
\label{bmlast}
\eeq
\beq
B_{g \go g}^{m,\H,{\rm MS}}(x,\mu_F)=\frac{5}{9N} f_g^{\H}(x,\mu_F)\,.
\eeq
Just as for the tagging functions the massive quark contributions to $A_{g \go g}$
and $B_{g \go g}$ are equivalent to a change of $n_f$ by one.  

It is easy to verify that the choice of the massive \mmsb
counterterms in Eq.~({\ref{eq:roneb}-\ref{eq:rfourb}) yields crossing
functions that are equal to those of the massless case 
(given in Eq. (3.39) and (3.40)
\footnote{We note a typo in Eq.~(3.40), 
the sign after the delta function in the expression for 
$B_{q \go q}^{\H,\msb}$ 
should be changed.}) in Ref.~\ref{GGK93}.  

In both factorization schemes the large logarithms 
$\ln(\Q/m)$ due to initial state collinear radiation are resummed
via the evolution of the parton distribution functions.

\subsubsection*{Matching at low $\Q$}

The PSS method, with the extensions presented in this paper,
provides a natural framework for the implementation of a
variable flavor number scheme (VFNS) 
\cite{ACOT94,Buza,LaiTung,MRRS,ThorneRoberts,Collins,Scalise} 
in the calculation 
of differential cross sections. 
The name refers to the fact that in such a 
scheme the number of active flavors depends on the value of \Q. 
Here we merely sketch the outlines of such an implementation,
we hope to provide a complete example
in the future \cite{KLnum}.
For a review of recent VFNS developments see 
e.g. Refs.~\cite{Collins,Smithconf}.

When the typical scale $\Q$  is close to the mass of the heavy quark, we
may use the PSS method extended by the soft gluon
radiation results of section 4.1.  There is no need to compute
analytically the contribution of (approximately) collinear
configurations involving the heavy quark, a numerical integration
should suffice. The heavy quark is not treated as a parton,
neither in the initial nor in the final state, i.e. it does not
contribute to the running of $\alpha_s$, nor to the evolution
of the parton distribution functions or fragmentation functions.  
To calculate the production cross section of a $D$ or $B$ meson, 
we have to convolute the parton-level results with a 
non-evolving fragmentation function (e.g. a
Peterson-Zerwas \cite{Peterson} function, or simply
$D_H^{\H}(z)=\delta(1-z)$). This calculational scheme, when
adopted irrespective of $\Q$, is called a fixed flavor number scheme (FFNS).

When $\Q$ is much larger than $m$, the fully
extended PSS method should be used, including the heavy quark crossing 
and tagging functions, factorized e.g in the $\mmsb$ scheme,
so that large logarithms $\ln(\Q/m)$ are resummed,
see earlier in this section.
Now the heavy quark is treated as
a parton, and it contributes to the evolution of 
$\alpha_s$, the parton distribution functions and fragmentation
functions. 

The issue is now how we should merge these two
descriptions into one calculational scheme.
For the discussion here we assume that the factorization
theorem holds even at a low scale close to the mass of the heavy quark,
see Ref.~\cite{Collins}. We take $\mu_R = \mu_F=\mu$ for
simplicity. At the end we shall put $\mu=\Q$.

Let us concentrate on the initial state. We
have in mind a cross section sufficiently inclusive to 
not require any fragmentation and tagging 
functions\footnote{E.g. one might
think of the deep-inelastic heavy-quark structure function.}.
The corresponding matrix element is computed with all partons 
in the final state, including the heavy quarks. 
Renormalization is performed in the CWZ scheme \cite{CWZ}, and all
soft and collinear singularities have cancelled due to the KLN \cite{KLN}
theorem. 

We start by choosing a matching scale, $\mu_0 \simeq m$, 
such that, for $\mu < \mu_0$, we use
$n_l$ active parton flavors in our calculation and 
treat the heavy quark as non-partonic,
while for $\mu > \mu_0$ we use $n_l+1$ active flavors
by treating also the heavy quark 
as partonic in the sense that it participates 
in the evolution of $\alpha_s$ and the parton 
distribution functions. 

For $\mu < \mu_0$ we cross, following Eq.~(\ref{eq:dshh}),
all possible ($n_l$) light partons to the 
initial state, and fold the result of each crossing with 
the corresponding $n_l$-flavor parton distributions functions.
We also add the contributions from the 
corresponding (\msb) light parton crossing functions. 
The evolution of the parton distribution
functions involves only the light degrees of freedom. 

For $\mu > \mu_0$ we replace all $n_l$-flavor parton 
distribution functions and $\alpha_s$ by their $n_l+1$ versions.
The relations between these $n_l$ and $n_l+1$ quantities
are given by so-called matching conditions,
computed for $\alpha_s$ to two-loop in Ref.~\cite{Kniehletal}
and for the parton distribution functions 
to one and two-loop in Ref.~\cite{CT86,Buza}.
We extend the sums over parton flavors and their 
crossings in Eq.~(\ref{eq:dshh})
to include the heavy quark. However, where we should add a heavy quark 
crossing function, we only need to add the corresponding counterterm 
function (e.g. those in Eqs.~({\ref{eq:roneb}-\ref{eq:rfourb})).
This is because, in the matrix element, 
all contributions, including near-collinear
ones, are computed numerically \footnote{For $\mu = \Q$ 
not too much larger than $\mu_0$,
neglecting terms of order $m^2/\smin$ when computing
heavy quark crossing and tagging functions, is not very accurate, 
due to the constraint $\smin \ll \Q^2$. Here it is therefore better
to compute contributions from the collinear region numerically.}. 
Of course, for $\mu \gg \mu_0$ one may 
use the heavy quark crossing functions derived earlier in this
section to compute the contributions of the collinear
region and counterterm functions.

To show that this calculational scheme constitutes
a VFNS, we must verify that both the low and high $\Q$ descriptions
are properly reproduced. Clearly, when $\mu \gg \mu_0$, we have an
$n_l+1$ flavor description, where the large logarithm $\ln(\Q/m)$,
converted by the counterterm functions into $\ln(\Q/\mu)$,
may be eliminated from the matrix element and resummed into
the parton distribution function by the choice $\mu=\Q$.

For $\mu$ just above $\mu_0$ the requirement is 
that the cross section computed in the VFNS is 
the same as computed in the FFNS\footnote{Note
that we consider one loop matching here, in which $\alpha_s$
and the parton distribution functions are continuous
across $\mu_0$, in the $\msb$ scheme.}.  
The argument essentially follows Ref.~\cite{ACOT94}.
For $\mu > \mu_0$, the difference with the FFNS cross section
is mainly due to the evolution of the gluon distribution function caused
by gluon splitting into a heavy quark pair,
and the counterterm functions.
For $\mu\;\gtap\; m$
(putting $\mu_0=m$, and
suppressing all irrelevant arguments),
the former contribution can be written as
\begin{equation}
  \label{eq:match}
d\sigma_{\rm NLO}^{(n_l+1)}({\rm VFNS}) - d\sigma_{\rm NLO}^{(n_l)}({\rm FFNS})
\simeq 
 \left(\alpha_s\, \left({N\over 2\pi}\right)
\ln\left({\mu\over m}\right)\, \frac{1}{2}P_{g\go q\bar{q}} 
\right)\, d\sigma_{\rm LO}^{(n_l)} 
\end{equation}
with $\mu$ not much larger than $m$. 
The corresponding counterterm functions (see e.g. Eq.~(\ref{rthree})),
precisely cancels the $O(\alpha_s)$ term in 
Eq.~(\ref{eq:match}), as required.

\section{Conclusions}

In this paper we extended the phase space slicing method
of Giele, Glover and Kosower \cite{GG92,GGK93} for computing NLO corrections to
jet cross sections to incorporate fragmentation functions 
and heavy quarks.  This makes the method applicable to any reaction 
in which a particular final state hadron (light or heavy) is tagged.

The extension to fragmentation functions for massless partons amounted to
a generalization of the crossing function approach \cite{GGK93} to 
the final state, and led us to introduce a new class of functions
we named tagging functions.

The extension to heavy quarks consisted of two parts. 
Effects of soft radiation off heavy quarks were straightforwardly included.
Contributions from collinear radiation involving heavy quarks were
included in heavy quark crossing and tagging functions.
We showed that heavy quark tagging functions led
naturally to the Mele-Nason perturbative fragmentation functions \cite{MN}.
We described briefly how our results might be included
in a variable flavor number scheme.

In this paper we developed the formalism of the extended PSS method.
Its practical use, and its application in the context of a variable
flavor number scheme, will be assessed in future work \cite{KLnum}.

\subsection*{Acknowledgments}
We would like to thank Walter Giele, Jack Smith and Wu-Ki Tung for 
useful discussions. E.L. would like to thank 
the Institute for Theoretical Physics in Stony Brook for its hospitality.
Part of this work was done while S.K was at Fermilab.
\appendix

\end{document}